# SoK: An Essential Guide For Using Malware Sandboxes In Security Applications: Challenges, Pitfalls, and Lessons Learned

Omar Alrawi, Miuyin Yong Wong, Athanasios Avgetidis, Kevin Valakuzhy, Boladji Vinny Adjibi,
Konstantinos Karakatsanis, Mustaque Ahamad, Doug Blough, Fabian Monrose, Manos Antonakakis
{alrawi, miuyinyong, avgetidis, kevinv, sadjibi3, karakatsanis, ma3, db209, fabian, manos}@gatech.edu
Georgia Institute of Technology

*Abstract*—Malware sandboxes provide many benefits for security applications, but they are complex. These complexities can overwhelm new users in different research areas and make it difficult to select, configure, and use sandboxes. Even worse, incorrectly using sandboxes can have a negative impact on security applications. In this paper, we address this knowledge gap by systematizing 84 representative papers for using x86/64 malware sandboxes in the academic literature. We propose a novel framework to simplify sandbox components and organize the literature to derive practical guidelines for using sandboxes. We evaluate the proposed guidelines systematically using three common security applications and demonstrate that the choice of different sandboxes can significantly impact the results. Specifically, our results show that the proposed guidelines improve the sandbox observable activities by at least $1.6x$ and up to $11.3x$. Furthermore, we observe a roughly $25\%$ improvement in accuracy, precision, and recall when using the guidelines to help with a malware family classification task. We conclude by affirming that there is no "silver bullet" sandbox deployment that generalizes, and we recommend that users apply our framework to define a scope for their analysis, a threat model, and derive context about how the sandbox artifacts will influence their intended use case. Finally, it is important that users document their experiment, limitations, and potential solutions for reproducibility.

## 1. Introduction

Malware sandbox systems are core tools in malware analysis and security applications, and they are integrated in intrusion detection systems [1], [2], forensic analysis pipelines [3]–[6], automated reverse engineering tools [7]–[9], and threat intelligence services [10]. However, correctly using sandboxes is challenging, mainly due to the plethora of choices in implementations, monitoring techniques, and analysis configurations, such as choosing between emulation, virtualization, or bare-metal. In fact, the choice of dynamic analysis and monitoring techniques can have a varying impact on the sandbox analysis results [11], [12], including missing or incorrect artifacts, sandbox evasion, misleading forensics results, and skewed behavior reports [13]. As a result, the security applications depending on those artifacts can be negatively impacted by false positive detections, inconclusive forensic analyses, and poor quality threat feeds.

Prior works that survey dynamic malware analysis techniques [11], [12] summarize different methods for researchers to build and improve sandboxes. Other works [14]–[16] survey evasion techniques or systematize binary instrumentation methods. The closest systematization from the user perspective for designing malware experiments [13] was published in 2012. This particular work provides applicable guidelines for malware experiments in general and does not specifically focus on sandboxes. Besides, since 2012, the computing environment, analysis methods, and malware threats have changed dramatically. There are new types of malware threats like ransomware and cryptominers; new analysis methods such as memory forensics and hardware tracing; and new computing environments such as cloud and IoT platforms. Moreover, prior systematization looks at designing malware experiments, which also includes using sandboxes, but does not provide the depth required to understand the implementations, monitoring techniques, and parameter configurations of sandboxes.

Despite the available literature on dynamic malware analysis, incorporating sandboxes into new security applications remains a challenging task. This is largely attributed to the complexity of sandbox deployments, which poses unique challenges to non-expert users. For example, researchers in interdisciplinary security fields may not have the system experience to consider sandboxes for their application or may be overwhelmed by the choice of technology such as hardware, hypervisors, operating systems (OS), and software services. In fact, because most of the literature surveyed lacks pertinent details about the sandbox deployment, it is very unlikely that users will reproduce the claimed results. Therefore, our objective is to bridge this knowledge gap by studying previous work *use* of sandboxes. We seek to derive a set of guidelines from prior works to help users effectively incorporate sandboxes in diverse security applications and avoid common pitfalls.

Unlike previous works [12]–[17], we study how researchers *use* sandboxes to understand the *practices* not only as a whole, but also how they can contribute to a systematic approach for incorporating sandboxes into new security applications. We achieve this goal by studying over 350 papers spanning two decades and systematizing 84 representative papers. Specifically, our work makes the following contributions:

- A component-based framework to simplify sandbox deployments and configuration for three classes of security applications, namely detection, observational studies, and anti-analysis.
- An application of the framework to systematize

84 papers on sandbox practices and derive seven guidelines to bridge the knowledge gap for sandbox end-users.
- A demonstration of how to apply the derived sandbox guidelines for a set of routine tasks, namely blocklist creation, malware behavior extraction, and malware family classification.
- An evaluation showing how the guidelines improve sandbox use by comparing the performance between two sandbox deployments, namely one that applies the proposed guidelines and another that does not.

Our systematization finds that using generic sandboxes can have limited coverage and can be highly skewed toward more prominent malware families. Conversely, applying advanced sandbox techniques is far from trivial, and even if end-users follow strict guidelines and best practices, there are no guarantees that the sandbox results will meet the desired outcome. In particular, the monitoring technique and the choice of analysis parameters appear to be the most important when using sandboxes. A transparent monitoring technique can improve the fidelity of sandboxes and therefore improve their analysis efficacy. Similarly, configuring a balanced analysis environment with signs of wear and tear, such as documents, images, videos, and browser activities, can potentially improve the analysis results. Ultimately, researchers should focus on defining the scope of their analysis, a threat model, and obtaining context about how sandbox artifacts will influence their application. Based on these observations, we propose seven guidelines for improving the use of sandboxes in security applications.

We evaluate the proposed guidelines systematically through three empirical experiments and demonstrate from the end-user perspective that the choice of different sandbox configurations can impact the desired outcomes. Specifically, we analyzed $1,471$ real malware samples that span eight different families to create a blocklist, extract malware behavior, and classify malware families. Our results show that the proposed guidelines can improve the observable artifacts of the sandbox by at least $1.6x$ and up to $11.3x$. Our family classification results show a 25% improvement in accuracy, precision, and recall when applying the guidelines. Moreover, by combining artifacts from both sandboxes, some improvement in classification results can be attained. These findings indicate that combining multiple guided sandbox deployments with different configurations can be an effective strategy. That said, even with a well-configured sandbox, we show that problems with failed execution, anti-analysis, and missing dependencies are not uncommon. We conclude with a set of recommendations and make our dataset and code available to the research community.

## 2. Framework and Methodology

Figure 1 provides an overview of the malware sandbox components, namely the implementations, the monitoring techniques, and the analysis parameters. The framework highlights each of these layers and their underlying configurations. We also differentiate between sandbox *end-users* and *developers/researchers*. In particular, sandbox developers and researchers build and improve the underlying sandbox implementations and monitoring techniques, while end-users configure and deploy sandboxes for dynamic malware analysis. End-users use sandboxes for a specific application, which we describe in the next section. In this work, we systematize the sandbox literature to distill guidelines for **configuration and analysis** methods for sandbox usage (dotted box).

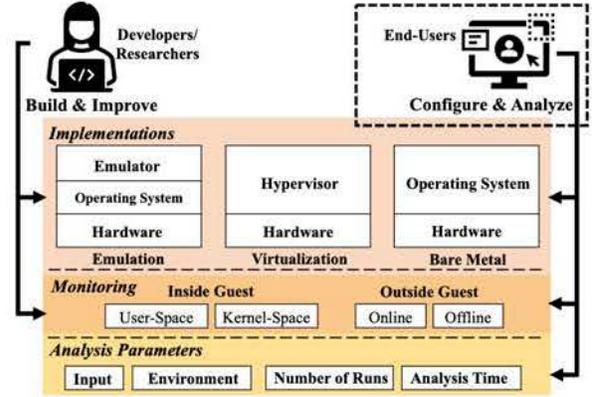

Figure 1: Overview of the systematization framework and stakeholders relationship. This work derives guidelines from the end-users perspective (dotted rectangle).

**Malware Sandbox Implementation.** In Figure 1, there are three types of *implementations*. The first type is **emulation**, which relies on implementing a virtual machine purely in software (i.e QEMU [18]). The second type is **virtualization**, which is a specialized operating system (OS) that segments resources and manages access to physical hardware for virtual machines (i.e. Xen [19]). The third type is **bare-metal** or physical machine, which runs an OS directly on the hardware. The key difference between the three types is how much of the physical resources (hardware) are implemented in software versus hardware. In modern computing, some pure software emulation systems use paravirtualization and virtual hardware modeling (virtio [20]) to improve performance, which overlaps with virtualization techniques. Figure 1 provides a **generic** model of sandbox implementations in the context of malware analysis, and we note that there can be many hybrid models.

**Malware Sandbox Monitoring.** In Figure 1, the *monitoring* section has four types of techniques that are built into the malware sandbox implementation or that researchers can extend for additional monitoring capabilities. These techniques include inside-guest user-space, inside-guest kernel-space, outside-guest online, and outside-guest offline monitoring. Inside-guest refers to monitoring tools that run inside the analysis environment, either in user-space as a program or in kernel-space as a driver, respectively. Outside-guest refers to monitoring tools outside the analysis environment, monitoring online (during analysis) such as VM introspection [21] or offline (post-analysis) such as forensic tools [22], respectively. Moreover, there is a myriad of monitoring capabilities [15] that researchers can develop inside or outside an analysis environment, depending on the type of sandbox implementation (emu-lation, virtualization, or bare metal). For additional information, Egele et al. [11] present an extensive survey of dynamic analysis monitoring approaches.

**Malware Sandbox Analysis Parameters.** In Figure 1, the *analysis parameters* section has four parameters that end-users can configure. The **input** parameter includes startup scripts, interaction scripts, and application instrumentation. These inputs help prime the analysis environment to trigger malware execution. For example, a startup script can create artifacts in the environment to show signs of wear and tear [23] or an interaction script can interact with system utilities to trigger rootkits [24]. The **environment** parameter includes the installation and configuration of the OS, system settings, user profiles, applications, and network settings. The **number of runs** parameter specifies how many times the sample is analyzed. Finally, the **analysis time** parameter includes static or adaptive analysis time. The time parameter determines the duration of the analysis, which can reveal dormant malware functions as the time parameter is varied. An advanced technique includes manipulation of the system's real-time clock to speed up time or report artificial time values, which can be achieved in an emulated and virtualized implementation.

**Sandbox End-Users.** As shown in Figure 1, we differentiate between sandbox developers/researchers and end-users. Sandbox developers and researchers must consider and make trade-offs between sandbox design properties, including transparency, scalability, extensibility, and isolation. For completeness, we provide a brief survey of the sandbox system design literature in the Appendix A. However, end-users generally do not need to consider these design decisions because they use sandboxes by configuring and analyzing malware [25]. In this paper, we systematize the sandbox literature from the perspective of *end-users*. Specifically, we group end-user sandbox applications into three categories, namely *detection*, *observational*, and *anti-analysis*. We provide more details on each of these categories in the next section.

## 2.1. Systematization Methodology

We systematize the literature on the *use* of sandboxes by examining their implementations, monitoring methods, and analysis parameters. We sourced more than 350 papers that span 20 years of research from top security conferences, journals, and workshops, then selected 84 papers that provide a comprehensive view of how security researchers (experts) have used sandboxes. We started by searching Google Scholar for dynamic malware sandbox analysis keywords, such as "malware," "dynamic analysis," and "sandbox." We identified the earliest works describing **dynamic malware sandbox analysis** and set them as our initial set.

We then searched for later works in two ways. First, we used the same set of keywords to search year-by-year starting from the earliest work up to the current year. Second, we manually went through later papers that cite earlier sandbox system papers and collected them. We refined our keyword set based on the newly identified papers and repeated the process until we enumerated all relevant papers. We selected 84 papers that are the most relevant by identifying how they uses sandboxes and their applications. We define the *use* of sandboxes as academic work that dynamically analyzes *malware* in an *x86/64-based sandbox system* for Microsoft Windows. We then prioritized the ranking of the paper based on relevance to a problem area (*novelty*), the number of citations (*impact*), and the publication venue (*visibility*).

## 3. Categories of Sandbox Usage

We categorize the *use* of sandboxes into detection papers, observational papers, and anti-analysis papers and group them in Table 1. It is important to note that these categories are non-exhaustive and that some papers will have dual applications. In such cases, we consider the primary goal of the paper as a classification discriminant. For example, Kim et al. [26] study obfuscation techniques (anti-analysis) in benign and malicious binaries to build a detector based on obfuscation features/tactics. Since the purpose of the work is to discriminate between benign and malicious programs, we classify the primary goal of the work as detection.

### 3.1. Detection

A common goal between detection papers is to identify if an unknown binary is malicious or benign. Detection papers use sandboxes to examine the behavior of unknown binaries through a monitoring component that provides input to the detection system. The detection papers surveyed examine dataset sizes of malware ranging from four samples to 150K samples. They rely on system and network artifacts as input for their detection techniques. From the 25 detection papers, 14 use system and network artifacts, six use only network artifacts, and 12 use only system artifacts. These variations account for different vantage points that detection systems use to identify malware, i.e., on-host versus off-host. In addition, they may have compute or visibility constraints (granularity of sandbox artifacts).

For example, Kirda et al. [27] propose an on-host in-browser analysis technique to detect malicious browser helper objects (BHO) whereas Antonakakis et al. [28] and Bilge et al. [1] propose detection techniques for internet service provider (ISP) networks (off-host). The first detection system requires a fine-grained system and browser behavior while the others only require DNS traffic (granular network artifacts), respectively. Furthermore, on-host detection can examine different granularity of activities such as CPU instructions [29], system calls [30], disk and memory forensics [31], [32], network activity [4], or a combination of all of them [33]. Network artifacts can range from full traffic capture [34] to flow [35] and even include protocol specific like DNS [1], [28] or HTTP [36]. *Detection systems rely on sandboxes to detect malware or evaluate their efficacy.* For example, Yen et al. [37] evaluate their system's detection robustness by using a small set of malware network traces and mixing them with predominately benign network traffic.

**Summary.** Detection papers have to deal with variability found in the choices of implementations, monitoring, and analysis parameters that can affect the reliability, visibility, and coverage of the sandbox results. Sandboxes can potentially have limited coverage, contaminated artifacts, and mislabeled behavior.

### 3.2. Observational

Observational papers analyze *known* malicious files. The dataset sizes for observational applications range from a single sample to millions, which varies based on the analysis depth and monitoring method. For unpacking malware (which uses instruction-level tracing), the dataset size ranges from 3 samples to 3,400, while the dataset size for longitudinal malware studies (which uses high-level behavior) ranges from 6,300 samples to 31M. The depth of analysis also relates to the artifacts studied (system or network). Of the 38 papers, 21 only use system artifacts, nine only use network artifacts, and eight use both. System artifacts have different levels of abstraction because they are dependent on the monitoring tool (system events, API, instructions, file system artifacts, etc.). Network artifacts have less variability because they can be observed from outside an analysis environment in a standardized form (i.e. *full packet capture*).

Of the 38 articles surveyed, we found five subcategories that measure or investigate malware threats using sandboxes. They deal with unpacking and deobfuscation, labeling and remediation, investigating *C&C* operation, automating reverse engineering, or studying malicious behavior over time (longitudinal). These subcategories use sandboxes in different ways. For example, some [38]–[41] look at labeling malware according to their behavior. Bailey et al. [38] and Paleari et al. [40] study how malware changes the state of a machine with different goals, namely identifying malware families or recovering from malware infections, respectively. *Observational studies use sandboxes very broadly.*

**Summary.** From a design perspective, extensibility and scalability are key properties seen in observational studies. End-users can choose sandbox monitoring capabilities for different depth of analysis and monitoring granularity. Scalability enables large-scale experiments in observational studies, which appear to favor suitable sandbox implementations like emulation and virtualization.

### 3.3. Anti-analysis

Malware authors use anti-analysis to evade sandboxes [14]. Anti-analysis papers try to make sandboxes more robust by detecting and thwarting anti-analysis methods. The surveyed anti-analysis papers use malware dataset sizes ranging from seven to 110K. They rely on system and network artifacts together or only system artifacts. Lindorfer et al. [42] use network artifacts to identify anti-analysis activities, namely, if malware queries external network resources for time, response strings, or a list of known sandbox IP addresses. Furthermore, network artifacts can help identify previously analyzed malware and avoid re-analyzing them [43]. For example, Bayer et al. [44] rely on both network system calls and network traffic traces observed on the wire.

Most malware in the wild use anti-analysis tactics [3], [42], [43], [45] to evade detection and reverse engineering. Specifically, anti-analysis papers deal with detecting anti-analysis samples [3], [9], [42], [45], studying anti-analysis tactics [46], [47], bypassing anti-analysis methods [48]–[51], and improving the resource utilization of sandboxes [43], [44], [52]. Fundamentally, anti-analysis papers improve sandboxes' transparency and scalability. For example, Willems et al. [49] propose a transparent and scalable monitoring technique using CPU features for anti-analysis malware.

**Summary.** In anti-analysis applications, end-users use variations of sandbox parameters to identify divergence, environment sensitivity, and alternate behavior that can serve as a baseline for their applications. Network artifacts appear to be more stable than system artifacts, and they can serve as an indicator for detecting anti-analysis behavior. However, in many cases, malware behavior is directly dependent on the malware operator's actions, which can change over time and across different environments.

## 4. Sandbox Applications and Usage

This section presents the systematization of sandbox practices for each application category. We distill these practices into a set of guidelines. These guidelines are not exclusive to each category, and end-users should use them to guide their experimental setups. We emphasize that some of the guidelines may appear to experts as common knowledge, but they are not documented with clear examples. Our approach is meant to document and disseminate sandbox practices across different areas of security, including interdisciplinary security research. For example, researchers in law and policy can apply empirical data to support cyber governance [53], military research in cyberspace [54], and cyber foreign policy [55]. Additional areas include management such as cyber insurance [56] and human-computer interaction (HCI) such as human factors in malware detection [57].

Table 1 lists the systematized literature by *Experiment Metadata*, *Implementation*, *Monitoring*, and *Analysis Parameters*. The experiment metadata documents the size of the dataset and the artifact types, namely system (*Sys*) or network (*Net*). The implementation section identifies the type of sandbox in use, namely emulation (*Emu.*), virtualization (*Virt.*), and bare-metal (*Metal*). The monitoring section documents how each paper implements its monitoring technique, namely inside-guest user-space or kernel space (*IGU/IGK*) and outside-guest online or offline monitoring (*OGOn/OGOff*). The online and offline approach refers to artifact collection during (online) or after (offline) the malware execution, respectively. The analysis parameters document the experiment setup for custom inputs, analysis environments (*Env.*), number of analyses (*Exec-n*), and analysis period (*Exec-t*) per paper.

### 4.1. Sandbox Implementation

**Detection Papers.** The role of a sandbox can influence the choice of implementation as shown in Table 1. We note that we found no evidence or correlation showing one implementation is better than another. Rafique et al. [34] is the only detection work that considers using more than one implementation (emulation and virtualization) to improve the chances of observing malicious network artifacts. These artifacts are obtained through a monitoring module that runs either inside or outside the analysis environment and each technique makes a trade-off between the four sandbox properties (see Appendix A). Lastly, we find that virtualization-based sandboxes appear to be favored for detection systems.

TABLE 1: Systematization using our framework across implementations, monitoring, and analysis parameters.

| App. | Paper | Experiment Metadata | | | Implementation | | | Monitoring | | | | Analysis Parameters | | | | Safety/Ethics Discussed |
|---|---|---|---|---|---|---|---|---|---|---|---|---|---|---|---|---|
| | | Size | Sys | Net | Emu. | Virt. | Metal | IGU | IGK | OGOn | OGOff | Input | Env. | Exec-n | Exec-t | |
| Detection | Kirda06 [27] | 33 | ✓ | | | | | ✓ | | | | ✓ | ✓ | | 50s | |
| | Li07 [31] | 1.4K | | | | ✓ | | | | ✓ | ✓ | ✓ | ✓ | 3 | 170s | ✓ |
| | Chris07 [58] | 16 | ✓ | | | ✓ | ✓ | ✓ | | ✓ | | ✓ | ✓ | 3 | 1m-4m | ✓ |
| | Litty08 [29] | 9 | ✓ | | | ✓ | | | ✓ | ✓ | | ✓ | | | | ✓ |
| | Marti08 [59] | 7 | ✓ | | ✓ | | | | | ✓ | | ✓ | | | | ✓ |
| | Rieck08 [30] | 10K | ✓ | | | ✓ | | ✓ | | | | | | | | |
| | Yen08 [37] | 4 | | ✓ | | ✓ | | | | ✓ | | | | 1 | 1h | |
| | Bayer09 [33] | 75K | | ✓ | ✓ | | | | | ✓ | | ✓ | | | | |
| | Anton10 [28] | | | ✓ | | | ✓ | | | | | | | | 5m | ✓ |
| | Fredr10 [60] | 912 | ✓ | | ✓ | | | ✓ | | ✓ | | | | | 2m | ✓ |
| | Perdi10 [36] | 25.7K | | ✓ | | | | | | ✓ | | | | | 5m | |
| | Jacob11 [4] | 37.5K | ✓ | ✓ | ✓ | | | ✓ | | ✓ | | | | | 4m | |
| | Ander11 [61] | 1.6K | ✓ | | | ✓ | | | | ✓ | | | | | 5m | |
| | Cui12 [32] | 150K | ✓ | | | | ✓ | | | ✓ | ✓ | | | 1 | 2m | |
| | Wille12 [62] | 7K | ✓ | | | ✓ | | | | ✓ | | ✓ | | | | |
| | Tegel12 [35] | 200 | ✓ | ✓ | | | | | | ✓ | | | | | | ✓ |
| | Palah13 [63] | 2.3K | ✓ | | | ✓ | | ✓ | | ✓ | | | | 1 | 2m | |
| | Rafiq13 [34] | 16K | ✓ | | | ✓ | ✓ | | | ✓ | | ✓ | ✓ | 2 | | ✓ |
| | Canza15 [64] | | ✓ | | | | | | ✓ | | | ✓ | | | 0.2s | ✓ |
| | Kharr16 [65] | 148K | ✓ | ✓ | | ✓ | | ✓ | | | | | ✓ | | 20m | ✓ |
| | Xu17 [66] | 320 | ✓ | | | ✓ | | ✓ | | ✓ | | | | 2 | 24s | |
| | Kim17 [26] | 100K | ✓ | | | ✓ | | ✓ | | ✓ | | ✓ | ✓ | | 1m | ✓ |
| | Sun20 [2] | 20K | ✓ | ✓ | | ✓ | | ✓ | ✓ | | | ✓ | | | | |
| | Zhang20 [67] | 1K | ✓ | | | | ✓ | | ✓ | ✓ | | | | | 1m | |
| | Count Summary | | 79% | 50% | 17% | 54% | 21% | 38% | 17% | 46% | 8% | 38% | 33% | AVG 1.24 | | 46% |
| Observational Studies | Royal06 [68] | 3.4K | ✓ | | | ✓ | | ✓ | | ✓ | | | | | 20m | |
| | AbuRa06 [69] | 192 | | ✓ | | ✓ | | | | ✓ | | | | | | ✓ |
| | Egele07 [17] | 21 | ✓ | | ✓ | | | | | ✓ | | ✓ | | | | |
| | Baile07 [38] | 8.2K | ✓ | | | ✓ | | | | ✓ | | | | | 5m | ✓ |
| | Kang07 [70] | 2.9K | ✓ | | ✓ | | | | ✓ | ✓ | | | | | 35s-70s | |
| | Polyc08 [71] | 307K | | ✓ | | ✓ | | | | ✓ | | | | | | |
| | Holz08 [72] | 5 | ✓ | ✓ | | ✓ | | ✓ | | ✓ | | | | | | |
| | Shari09 [73] | 16 | ✓ | | ✓ | | | | | ✓ | ✓ | | | | | |
| | Cabal09 [7] | 1 | ✓ | ✓ | ✓ | | | | | ✓ | | | | 1 | | ✓ |
| | Stone09 [74] | | ✓ | | ✓ | ✓ | | ✓ | | ✓ | ✓ | | ✓ | | | |
| | Quist09 [75] | 3 | ✓ | | ✓ | ✓ | | | | ✓ | | | | | 5m-12h | ✓ |
| | Bayer09 [76] | 900K | ✓ | | ✓ | | | | | ✓ | | | | | 4m | |
| | Holz09 [77] | 2K | ✓ | ✓ | | ✓ | | ✓ | | ✓ | | ✓ | | | >1m | |
| | Kolbi10 [39] | 6 | ✓ | ✓ | | ✓ | | | | ✓ | ✓ | | | 1 | 30m | ✓ |
| | Palea10 [40] | 200 | ✓ | | ✓ | | | | | ✓ | | | | 15 | | |
| | Krish10 [78] | 10 | ✓ | | | ✓ | | | ✓ | ✓ | | ✓ | ✓ | | 7d | |
| | Compa10 [79] | 11K | ✓ | ✓ | ✓ | | | | | ✓ | | | | 1 | | |
| | Leita10 [80] | 6.3K | ✓ | ✓ | | | | | | ✓ | | | | 1 | 4m | |
| | Nadji11 [81] | 2K | ✓ | ✓ | ✓ | | | | | ✓ | | | ✓ | 7 | 5m | ✓ |
| | Neugs11 [82] | 8.3K | ✓ | ✓ | ✓ | | | | | ✓ | | | ✓ | | 6m | ✓ |
| | Cabal11 [83] | 9.1K | | ✓ | | ✓ | | | | ✓ | | ✓ | | 2 | | ✓ |
| | Rosso12 [84] | 243K | | ✓ | | ✓ | | | | ✓ | | | | | 1h | ✓ |
| | Lindo12 [85] | 381 | ✓ | | ✓ | | | | | ✓ | ✓ | | | 33-88 | 2m | ✓ |
| | Hu13 [41] | 5.6K | ✓ | | | ✓ | | ✓ | | | | | | | 2m | |
| | Jagpa15 [86] | 100K | ✓ | | | ✓ | | | | ✓ | | ✓ | ✓ | | | |
| | Kharr15 [87] | 4K | | ✓ | | ✓ | | | ✓ | ✓ | | ✓ | | 1 | 45m | ✓ |
| | Yadeg15 [88] | 5 | ✓ | | | ✓ | | | | ✓ | | | | 3 | | |
| | Ugart15 [8] | 7K | ✓ | | | ✓ | | | | ✓ | | | | | 30m | |
| | Thoma15 [89] | 25M | ✓ | | | ✓ | | ✓ | | ✓ | | ✓ | | | | |
| | Han15 [90] | 1M | | ✓ | ✓ | | | | | ✓ | | | | | | |
| | Ugart16 [91] | 3 | ✓ | | | ✓ | | | | ✓ | | ✓ | | 3 | | |
| | Thoma16 [92] | 1.47M | ✓ | | | ✓ | | ✓ | | ✓ | | ✓ | ✓ | | | |
| | Lever17 [93] | 26M | | ✓ | | ✓ | | | | ✓ | | ✓ | | | >5m | |
| | Farin17 [94] | 19K | ✓ | | | ✓ | | ✓ | | ✓ | | ✓ | ✓ | 8 | 35m | ✓ |
| | Cozzi18 [5] | 10.5K | ✓ | | ✓ | | | | ✓ | ✓ | | | | | 5m | ✓ |
| | Haq18 [95] | 7.8K | ✓ | | ✓ | | | | | ✓ | | | | | 10m | |
| | Barr21 [10] | 31.8M | ✓ | | | ✓ | | ✓ | | ✓ | | | | | | |
| | Count Summary | | 76% | 46% | 46% | 57% | 0% | 27% | 8% | 78% | 16% | 30% | 27% | AVG 2.68 | | 35% |
| Anti-Analysis | Chen08 [47] | 7K | ✓ | | | ✓ | ✓ | | ✓ | | | ✓ | | 3 | 2m | |
| | Balza10 [3] | 10 | ✓ | | ✓ | | | ✓ | | ✓ | ✓ | | | 2 | | |
| | Round10 [48] | 200 | ✓ | | | ✓ | | ✓ | | ✓ | | | | | | |
| | Bayer10 [44] | 11K | ✓ | ✓ | ✓ | | | | | ✓ | | | | 3 | 20s-300s | |
| | Lindo11 [42] | 1.8K | ✓ | ✓ | ✓ | | | | ✓ | ✓ | | | ✓ | 12 | | |
| | Kawak13 [50] | 7 | ✓ | | ✓ | | | | | ✓ | | | | 3 | 5m | |
| | Kirat14 [45] | 110K | ✓ | | ✓ | ✓ | ✓ | ✓ | | ✓ | ✓ | | ✓ | 4 | | |
| | Xu14 [51] | 1.4K | ✓ | | | ✓ | | ✓ | | ✓ | | ✓ | ✓ | | 5m | |
| | Kirat15 [9] | 3.1K | ✓ | | | ✓ | | | | ✓ | | | | 2 | | |
| | Vadre16 [43] | 1.6M | | ✓ | | ✓ | | | | ✓ | | | | | 6m | |
| | Kuchl21 [52] | 84K | ✓ | | | ✓ | | ✓ | | ✓ | | ✓ | | | 15m | |
| | Count Summary | | 912% | 361% | 642% | 649% | 185% | 361% | 185% | 825% | 185% | 273% | 4538% | AVG 2.85 | | 0% |

**Observational Papers.** Interestingly, all surveyed works rely on emulated or virtualized sandboxes. For applications that use system-based artifacts, such as unpacking and automated reverse engineering, emulated and virtualized implementations are more flexible and easier to extend. For example, Caballero et al. [7] make use of in-memory data taint-flow analysis built on top of an emulated sandbox to automatically reverse engineer custom malware protocols. Such implementation is more involved on a bare-metal system. For network-based artifacts, emulated and virtualized implementations provide better scalability and resource utilization than bare-metal sandboxes. From a scalability approach, Barr-Smith et al. [10] analyze over 31M samples using virtualization-based sandbox to understand how malware abuses system services and utilities. From a longitudinal approach, Bayer et al. [76] study malware using an emulation-based sandbox, Anubis, which analyzed 900K samples over five years. However, in those five years, Anubis has gone through several feature enhancements that affect the sandbox reports, which create inconsistent results due to missing artifacts. A drawback for emulated and virtualized sandboxes is that the hardware transparency is imperfect and has implications with regards to anti-analysis [14]. Nevertheless, it appears that observational studies make a trade-off between hardware transparency and scalability and extensibility by favoring emulated and virtualized sandboxes.

**Anti-analysis Papers.** One of the unique aspects of anti-analysis papers is that they utilize multiple implementations *concurrently*. This approach compromises between transparency, scalability, and extensibility to provide better analysis coverage. Concurrent analysis curbs the time and location variability. Time variability refers to analyzing a stale malware sample versus a fresh sample with an active C&C server. Location variability refers to analyzing malware samples from the same geographical location of the sandbox's network. Controlling for time and location can potentially identify malware behavior divergence due to the sandbox implementation [96].

For example, Chen et al. [47] and Kirat et al. [45] use multiple sandbox implementations to study anti-analysis behavior. The works detect anti-analysis and identify evasion tactics [9] using sandbox systems like Ether [97], Cuckoo [98], and Anubis [99]. Anti-analysis papers provide working examples for end-users to deploy a robust sandbox that can potentially overcome many of the anti-analysis tactics found in the wild. However, deploying such a sandbox is far from trivial and requires heavy engineering efforts. Moreover, because malware authors are always improving their tactics, sandboxes should be maintained to deal with new anti-analysis tactics. Lastly, given the adversarial nature of anti-analysis applications, researchers appear to prioritize extensibility, which emulated and virtualized sandboxes offer.

**Takeaway.** Prior works appear to prioritize isolation, extensibility, and scalability over *hardware* transparency. The extensibility property can mitigate some of the hardware transparency issues by patching system calls to deceive malware. However, malware can use side channels, like timers, to identify hardware implementation inconsistencies. Although bare-metal sandboxes have perfect hardware transparency, they do not appear to be widely used, as seen in Table 1.

### 4.2. Monitoring Techniques

**Detection Papers.** Table 1 documents the monitoring components in use by detection papers. Monitoring techniques impact the *transparency* and *scalability* of a sandbox. Earlier papers rely on inside-guest user-space and kernel-space components for behavior monitoring because they are easier to implement and provide behavior context. For example, Jacob et al. [4] use inside-guest hooking for the Winsock APIs because it provides data flow context for the *send* and *recv* system calls. On the other hand, monitoring techniques inside the analysis environment can violate the transparency requirements and make it easier for malware to evade.

Fredrikson et al. [60] rely on an inside-guest user-space monitoring tool, similar to *strace*, to collect system calls and note that malware can thwart their monitoring tool. However, Anderson et al. [61] address this problem by using Ether [97], which uses an outside-guest monitoring component. Another approach is to use inside-guest kernel monitoring, which provides better transparency and it is easier to extend to bare-metal systems [67]. Outside-guest approaches can include network traffic monitoring or use memory introspection as shown by several works [33], [59], [61]. Anubis [99] uses outside-guest monitoring but requires environment modification to log system calls [4]. In addition, Sun et al. [2] combine two monitoring techniques, namely inside-guest (API hooking) and outside-guest online, to detect malware DNS behavior.

Monitoring techniques also have a timeliness property that detection papers can customize based on their needs. Timeliness refers to whether detection is in real-time or post-analysis. For a post-analysis example, Cui et al. [32] use an outside-guest offline approach by inspecting crash dump memory images to detect malware. For a real-time example, Canzanese et al. [64] use the first 200ms of analysis to detect malware. Detection papers leverage varying monitoring techniques to address challenges such as contextualizing behavior and balancing transparency and scalability. Using expert rules like *Sigma* and *Yara* to contextualize behavior artifacts should be carefully considered since malware behavior changes across time [96]. In addition, the variations of monitoring techniques can create a split-view in the observed malware behavior [4], [27], [30], [58], [60]. For example, a network trace and a system call trace provide different levels of artifact context such as event order and dependency.

**Observational Papers.** Table 1 documents the different monitoring techniques in use by observation papers. Most papers rely on monitoring components that reside outside the analysis environment with some exceptions. For example, unpacking [68] with debuggers, tracking browser extension abuse [86], [89], [92], capturing system call sequences [41], and tracking malware operations [72], [74], [77], [94] rely on inside-guest user-space monitoring. The only factor that seems to influence the use of inside or outside monitoring components is the default monitoring implementation available in a sandbox. For example, several works [72], [77] use CWSandbox [100], which implements an inside-guest user-space Windows API hooking. Other works [86], [89], [92] build a customized sandbox to study browser extensions. These approaches help create consistency for individual studies, but other researchers cannot reproduce or verify the results.

Moreover, the implementation of inside-guest monitoring can vary from one sandbox to another. For example, the monitoring of Windows APIs (memory, file system, network), the hooking techniques (injection, instrumentation, direct memory modification), and the logging of artifacts (function names, parameters, and return values) can create split-views of malware behavior. Kernel-space implementations may offer more transparent monitoring that provides an encompassing view of malware behavior. For example, Kang et al. [70], Kharraz et al. [87], and Cozzi et al. [5] use inside-guest kernel monitoring to unpack malware, study ransomware, and study Linux malware, respectively. Transparency is also influenced by the monitoring implementations, which anti-analysis malware can potentially detect and hide their malicious behavior.

Outside-guest monitoring can improve sandbox transparency. Ugarte-Pedrero et al. [8] use outside-guest system call tracing to study unpackers through system artifacts. Furthermore, network artifact collection should always prioritize outside-guest monitoring to be transparent. Outside-guest network monitoring can extract IRC $C\&C$ servers [69], peer-to-peer encryption keys [72], and malware hosting infrastructure [13], [83]. Although network captures lack the system context of the behavior (network-to-process id), they can provide a consistent view of malware behavior. For example, Polychronakis et al. [71] monitor the network behavior of malware in a sandbox to study how malware interacts with the $C\&C$ server and present a network view of the interaction. In general, outside-guest monitoring tools appear to be the most popular for collecting network artifacts.

**Anti-analysis Papers.** Anti-analysis papers often rely on multiple monitoring components with tailored binary analysis to detect and overcome anti-analysis malware. For example, Balzarotti et al. [3] extend the capabilities of the Anubis sandbox to support record-and-replay of API parameters and return values using two different systems (a reference and an experiment system). Most anti-analysis papers use outside-guest monitoring methods to capture malware behavior that relies on emulation- and virtualization-based sandboxes. A popular approach to detecting and bypassing anti-analysis malware is by incorporating bare-metal-based sandboxes to provide a transparent hardware environment, but their monitoring techniques can be limited, and in some cases, they can be evaded by malware [101]. These limitations give the users a trade-off between transparency and extensibility. Applications seem to compromise between the two choices by leveraging binary instrumentation to collect behavior profiles [48], [51] or rely on a customized kernel module to monitor behavior [3], [42] for inside-guest monitoring.

**Takeaway.** Outside-guest online monitoring appears to be the most common approach, while outside-guest offline is the least common. Detection papers often rely on multiple monitoring techniques and make use of inside-guest monitoring more often. However, within inside-guest, user-space monitoring tools are more common across all application categories. Anti-analysis malware can detect and evade monitoring tools by circumventing hooking and delaying execution.

### 4.3. Analysis Parameters

**Detection Papers.** A third of the detection papers specify input for the sandbox analysis. They rely on both generic and custom input. For example, a custom input can induce activities by opening a specific website in a browser [27] or opening a specific version of a document reader to trigger an exploit [62]. Generic input can simulate system activities such as mouse movements [34], [67], clicking windows [26], or user browsing [60]. In addition, input can be applied to the network application to simulate internet traffic [58], modify traffic [59], simulate network services [33], or redirect traffic to a honeypot [64].

A third of the detection papers customize their environment by installing software to trigger malware behavior such as browsers [27] or document readers (PDF) [31], [66]. Environment customization can entail populating a user profile to create an environment that looks similar to a real user's system to lure malware ransomware [65]. For instance, Willems et al. [62] use multiple environments with different versions of Adobe Reader to detect malicious PDF files. The environment customization can increase the chances of triggering malicious behavior, which can provide better coverage. Conversely, Zhang et al. [67] customize the system environment to contain sandbox artifacts that malware tends to evade to thwart malware infections. A fourth of the detection papers consider analyzing malware multiple times [31], [58]. However, the majority of papers either do not explicitly state the number of executions or we assume it is at least once. The number of analyses is associated with other parameters such as different environments or analysis time (Exec-t). For example, Xu et al. [66] run suspicious PDF documents in two environments, namely macOS and Windows, to measure the divergence in the program behavior, which requires the sample to run twice.

Christodorescu et al. [58] run malware samples three times with varying analysis times (1 min, 2 min, and 4 min) to identify dormant behavior. Yet, the time parameter seems to be less related to detection requirements and more subjectively chosen. Several works [4], [28], [32], [36], [60], [63] choose a time interval between two minutes and six minutes with no direct justification. However, the time range among detection papers can range from very short (milliseconds) to very long (an hour). Canzanese et al. [64] detect and terminate suspicious applications in a production environment by running malware for as short as 200 ms. On the other hand, Yen et al. [37] collect malware network traces for one hour based on the assumption that an hour provides sufficient time for malware to communicate with the C&C server. Recent work by Kuchler et al. [52] systematically experiments with the time parameter and suggests that a two-minute time interval is sufficient for most sandbox applications.

**Observational Papers.** A third of the observational papers surveyed use a sandbox with custom input. These inputs simulate user activities to create an event [77], [87], trigger malicious browser extensions [17], [86], track ad injection abuse [89], attract malware operators [94], or study pay-per-install malware [92]. Furthermore, some papers specify input through network service simulation [71], [84] and network traffic manipulation [83] to extract key indicators for their study. Since some malware may have a stale $C\&C$ infrastructure, providing network inputs can be an effective approach to analyzing malware in a sandbox. Many of the works have a priori knowledge of the type of malware under analysis, which helps end-users craft the right inputs for the sandbox. However, a priori insights require end-users to manually analyze or reverse engineer the malware under investigation.

A fourth of the observational papers customize their environment. Farinholt et al. [94] study the tactics of malware operators using eight unique environments with different personas (default install, gamer, doctor, politician, academic, student, bitcoin miner, and teller) to lure malware operators to interact with an infected host. Additionally, Paleari et al. [40] use multiple analyses, running each sample 15 times, to identify divergence in behavior under different environments (locale, timezone, installed software) and to generate a generic remediation model for post-infection cleanup. Environment customization plays an important role in studying malware, but surprisingly it is rarely utilized by the surveyed papers. Environment artifacts give the analysis environment a sense of wear-and-tear [23] that some malware will check before revealing its intended behavior. In fact, Vadrevu et al. [43] report that as high as 62% of the malware samples they study (1.6M samples) lack sandbox activities, which can potentially be related to environmental factors.

Less than 20% of the observational papers execute the malware more than once. Observational papers use multi-analysis to identify anti-vm behavior [83] and analyze diverging behavior in different environments [40], [81], [94]. For example, Nadji et al. [81] analyze malware samples multiple times with varying network rules to identify alternative malware communication channels. Some observational applications require executing the malware multiple times to incrementally learn about code constraints to discover its behavior. For instance, unpacking papers [88], [91] run malware multiple times to identify control dependencies in layered obfuscation.

Additionally, the analysis period in a sandbox can affect the results. We observe a wide range of execution times varying from 35 seconds to seven days but most of the papers analyze malware an average of five minutes. Krishnan et al. [78] use a sandbox to evaluate their provenance-tracking system where they infect an environment for seven days, which is a less common use case. Lindorfer et al. [85] present a novel approach that simulates a long analysis period (three months) using save-and-restore analysis. They execute several samples for two minutes then snapshot the sandbox's execution state to be restored and run again the next day. They repeat the analysis between 33 and 88 days to study malware code evolution through repetitive but finite execution.

Other works analyze malware until an event is observed, such as $C\&C$ server communication [69], decryption key exchange [72], download of an executable [83], or any other event-based observation [86], [89], [92]. Some hybrid approaches combine event-based and time threshold parameters such as the work of Ugarte-Pedrero et al. [8], which defines a maximum time-out threshold of 30 minutes with rules for adaptive execution time.

**Anti-analysis Papers.** Similar to observational papers, anti-analysis papers retroactively use custom input. For example, Balzarotti et al. [3] and Xu et al. [51] collect API calls, API parameters, and function return values then use them as input for subsequent analyses. These works differ in their goal, Balzarotti et al. [3] detects anti-analysis malware whereas Xu et al. [51] bypass anti-analysis to capture the malware behavior. Furthermore, both works utilize multi-analysis environments to observe how malware behaves in different settings. Xu et al. [51] dynamically adapt the environment by running multiple analyses in parallel and selectively terminating instances that do not reach the target malicious code.

Other works aim to improve a sandbox's resource utilization. They do this by identifying polymorphic malware samples that have the same initial behavior and then terminate the sandbox analysis early. Bayer et al. [44] use generic behavior profiles generated from historic malware reports to selectively analyze samples that are different from previously analyzed malware. Similarly, Vadrevu et al. [43] propose an optimization for bare-metal sandboxes that only relies on network traffic to prioritize new malware samples. On the other hand, Roundy et al. [48] propose a selective hybrid analysis approach that combines static and dynamic analysis to optimize binary instrumentation for sandboxes. Furthermore, multiple analyses are sometimes associated with a customized environment. Lindorfer et al. [42] perform 12 parallel analyses using a variation of two monitoring techniques (inside- and outside-guest) and three different custom environments (software, user privilege, and locale). Their work combines variations from the monitoring and analysis parameter component seen in Figure 1 for a multi-analysis approach. Similarly, Kirat et al. [45] use one custom environment that contains saved credentials, browser history, and user document files but vary the sandbox implementations (emulation, virtualization, and bare-metal).

It appears to be a common practice for anti-analysis applications to analyze malware more than once since they require varying one or more features in each component of the sandbox. However, there are exceptions where researchers evaluate new sandbox technology (i.e. processor-level monitoring) using anti-analysis malware to demonstrate its efficacy [49]. Similar to detection applications, the execution time for the anti-analysis papers is between 20 seconds to 10 minutes. Overall, anti-analysis papers demonstrate promising use of sandboxes for robust and high-fidelity results but can require significant engineering efforts.

**Takeaway.** We find that less than 40% of the papers document their analysis parameters. Unfortunately, this practice creates a reproducibility and a knowledge gap. Analysis parameters appear to be one of the most important components for sandboxes and can greatly improve the chances of observing malware behavior. Moreover, the analysis parameters have a large array of configurations that end-users should narrow down by understanding how sandbox artifacts are used in their application.

### 4.4. From Practices To Guidelines

Our systematization finds that sandbox deployments are complex and should be carefully incorporated into security applications. End-users should **not** depend on generic sandboxes since their implementations, configuration, and analysis parameters may negatively impact outcomes. For example, the results may be skewed towards more prominent malware families [102]. Overall, there is no "silver bullet" practice for sandbox deployments, and applying advanced sandbox techniques is far from trivial. To help end-users navigate some of the complexities of using sandboxes, we propose several guidelines based on our in-depth analysis of the literature.

**Guideline-1: Simplifying sandbox deployments.** Compartmentalize components into manageable and configurable components. Simplify deployments by independently configuring each component. Use outside-guest offline monitoring techniques to simplify deployments. Outside-guest offline monitoring tools are decoupled from the implementation and analysis environment and can potentially provide sufficient analysis context.

**Guideline-2: Understanding sandbox trade-offs.** Use emulated and virtualized sandboxes as they offer better extensibility and scalability. Bare-metal sandboxes have perfect hardware transparency, but can be difficult to deploy for common security tasks.

**Guideline-3: Improving monitoring transparency.** Masquerade monitoring tools inside the analysis environment by randomizing their names and installation paths. Collect network traffic from outside the analysis environment. Minimizing the forensic tools inside the analysis environment can potentially thwart common malware evasion tactics.

**Guideline-4: Improving environment transparency.** Use common triggers in the analysis environment such as generic mouse/user interaction, software installation, and randomize user profiles. Use context about known malware to tailor the environment. Simulate internet services for stale or missing C&C servers. Iteratively analyze evasive malware to identify and address factors that contribute to the anti-analysis tactics.

**Guideline-5: Testing sandbox deployment.** Evaluate the environment configuration using malware with known behavior and contrast observed versus expected behavior. Test a sample of malware corpus on two different environments (highly configured vs. no configuration) to document the impact on observed behavior. Identify the environmental factors that inhibit the behavior of malware (i.e. missing software dependencies) and address them adequately.

**Guideline-6: Evaluating sandbox results.** Collect artifacts from at least two monitoring components to compare. Evaluate the correctness of the sandbox artifacts by referencing the network artifacts and activities. Use a reference and experimental sandbox deployment to identify anti-analysis behavior quickly. Normalize the results of the sandbox from different versions of the sandbox.

**Guideline-7: Optimizing sandbox analysis.** Analyze malware for two minutes initially, but also test other thresholds. Configure event-based triggers and timeout thresholds to end the sandbox analysis when malware crashes or desired event is observed. Simulate long-lived malware executions by snapshotting and saving the analysis environment to restore/resume the malware's execution later. Use network artifacts, such as the malware's C&C server, to identify polymorphic samples to avoid reanalyzing the same malware.

**Guideline-8: Safety and Ethics.** Ensure your planned experiment adheres to research community ethical norms and aligns with ethical considerations outlined by the security community [103], [104]. Seek approval from your institute or administrator point of contact before conducting malware experiments. Ensure the malware analysis system is adequately isolated from real systems and networks and adheres to administration policy. Configure the minimum network services in the malware analysis network to avoid unintended malware attacks. Employ traffic shaping techniques, such as traffic redirection and reflection, to limit the impact of malware analysis on the internet.

## 5. Evaluation: Sandbox Guidelines

In this section, we evaluate the proposed guidelines systematically through a set of empirical experiments. Our goal is to demonstrate from the end-user perspective that the choice of different sandbox configurations can affect the resulting findings. We refer to the work of Yong Wong et al. [25] for an inside look at the practice of malware analysis by security experts. Specifically, we use their framework to evaluate our proposed guidelines based on three types of tasks.

1) **Objective 1:** Create a blocklist based on sample hash, IP addresses, and domains (Tier 1-Task 1).
2) **Objective 2:** Extract malware behavior based on network and system artifacts (domains, IP addresses, files, registries, and processes) (Tier 2-Task 1).
3) **Objective 3:** Assign a malware family label based off the extracted network and system artifacts (Tier 2-Task 2).

Yong Wong et al. also highlight additional tasks such as generating reports and tracking tactics, techniques, and procedures (TTP) [105]. We exclude those tasks from the evaluation because they require a human-in-the-loop to assess the sandbox results. On the other hand, the three objectives can be measured empirically (either observed or not observed), which excludes any human bias. Therefore, we use the three objectives to demonstrate how our proposed guidelines can help end-users improve their use of sandboxes for the prescribed tasks.

### 5.1. Overview

Our empirical experiment takes the following approach. First, we configure and deploy a sandbox based on the guidelines proposed in the systematization, which we refer to as a guided sandbox (G). Second, we deploy another sandbox that runs with default settings and minimal required configurations (networking, disable antivirus, create user profile, etc.), which we refer to as an unguided sandbox (U). We then collect 1,471 recent malware samples spanning eight popular malware families and build ground truth signatures based on leaked source code, open source code, and manual malware analysis. The ground truth is a set of signatures for file, registry, process, and network artifacts that each family is expected to generate when executed in a sandbox. We then evaluate our signatures for a *benign* binary that we run in both sandboxes and ensure that our signatures do *not* match against any of the artifacts produced by the benign sandbox analysis (identify false positives). We then execute each malware sample in both sandboxes and observe the behavior across the four categories of artifacts (file, registry, process, and network). Based on the observed matches, we empirically quantify the execution of each malware family in both sandboxes in the context of the three objectives mentioned earlier.

### 5.2. Guided Sandbox Setup

We apply *guideline-1* and use our framework in Figure 1 to choose the implementation technology, monitoring technique, and analysis parameters.

**Guideline-2: Implementation Trade-offs.**

1) We use virtualized sandboxes because they offer better extensibility and scalability.

**Guideline-3: Monitoring Transparency.**

1) We minimize the forensic tools inside the analysis environment and collect network traffic outside the analysis environment using Second-Level Address Translation (SLAT) technology.

**Guideline-4: Environment Transparency.**

1) We gather context on the nature of malware families from public reports to install the software that the malware targets.
2) We implement a user-interaction script to induce mouse and keyboard clicks. We install 18 different software for file transfer (WinSCP, FileZilla, etc.), mail clients (OperaMail, Thunderbird, etc.), and instant messengers (ICQ, Skype, etc.). Table 6 in Appendix A contains the complete list.
3) We populate the analysis environment with 280 fake documents across the Desktop, Documents, Music, Videos, and Pictures. We include activity artifacts (saved passwords, cookies, etc.) for browsers, mail clients, instant messengers, and file transfer applications.

**Guideline-5: Test Sandbox Deployment.**

1) We manually analyze a random sample of malware families to understand their expected behavior. We use this information to test our deployment for malware execution.
2) We evaluated our malware corpus on two sandbox deployments, namely guided (G) and unguided (U) and measured the observed artifacts.

**Guideline-6: Evaluate Results.**

1) We evaluate both inside-guest user-space hooking and outside-guest introspection. We use Intel Extended Page Tables (EPT), for outside-guest online monitoring.
2) We inspect the network, file, registry, and process artifacts to identify malware execution.

**Guideline-7: Optimize Analysis.**

1) We configure a two-minute timeout for the analysis. However, we also apply an anti-evasion approach that identifies large parameter values for the *sleep* system call and modify it to smaller values for both sandboxes.
2) We terminate the analysis if the malware execution crashes or the two-minute timeout elapses to optimize the compute resources.

**Guideline-8: Safety and Ethics.**

1) We conducted our experiment in a dedicated malware network that supports isolation, minimal services (DHCP and DNS), and traffic shaping (forward, drop, rewrite, redirect, and reflect) to prevent attacks such as spam and DDoS.

We deploy the unguided sandbox with minimal configuration to allow us to execute malware. For example, we disable the anti-virus engine (Windows Defender) inside the guest to avoid the samples from being quarantined. All non-essential configuration parameters are left untouched.

### 5.3. Malware Dataset

**Malware Family Choice.** We use three sources of threat intelligence to identify actively used malware families, namely Malware Bazaar [106], CyberCrime tracker [107], and Any.run's malware trends [108]. At the time of writing, the five most popular malware families are *AgentTesla*, *Formbook*, *Lokibot*, *Ponystealer*, and *AZORult*. We prioritized malware families based on the availability of source code. We found leaked source code for *Lokibot*, *Ponystealer*, and *AZORult*. During our search for source code, we found additional source code for malware families that also appear to have been popular in the past, such as *Amadey*, *Neutrino*, and *BlackNET*. Although these families are not the most popular malware found in the wild today, we found that they are still in use and included them for a more complete evaluation.

**Malware Collection.** We sourced the malware binaries from VirusTotal using a combination of AV labels, Yara rules [109], Sigma rules [110], and IDS rules [111]. We use AVClass [112] for malware family labeling. We use strict criteria where all four indicators (AV, Sigma, IDS, and Yara) must agree on the malware family. We collected the most recent samples seen in VirusTotal (the last two weeks) and limited them to a maximum of 200 binaries per family. In the case of *Neutrino*, *Amadey*, and *Azorult*, this maximum could not be reached because our selective criteria of four indicators did not match 200 binaries from VirusTotal. Furthermore, we sampled 15 binaries per family and manually analyzed them statically and dynamically to ensure that the malware families are correctly classified. In total, our malware dataset consists of 1,471 malware samples spanning eight different malware families.

### 5.4. Methodology

**Malware Code Execution Order.** By inspecting the source code available for some families and manually analyzing others, we identified particular malware patterns during execution. More specifically, Figure 2 is an illustration of the general behavior observed for a malware binary during execution. We find that malware will attempt to persist on the system by moving the malware binary from the original execution path (i.e. Desktop or Downloads folder) and placing itself in the user profile directory (i.e. Users/<User Name>/). Then the malware will create or modify the system's registry to create a service or add itself to the startup entry to execute on system startup. These activities generate observable artifacts during execution in the form of file and registry creation, modification, and deletion. Figure 2 depicts this process by mapping the malware execution lifecycle (orange box) to the observed artifacts in a sandbox (green).

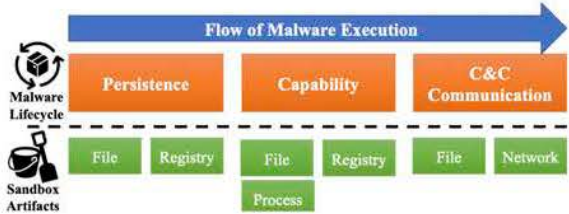

Figure 2: Typical execution flow of malware.

Similarly, we observe that the capability deployment phase of a malware (stealing credentials, dropping malware, privilege escalation, etc.) creates observable sandbox artifacts such as file, registry, and process (mutex) creation, modification, and deletion. Eventually, malware will attempt to establish communication with the C&C server to send data or receive commands. This phase adds an additional category of sandbox artifacts, namely the network communication (domains and IP addresses). By adapting a MITRE ATT&CK-like [105], [113] framework, we can generalize the observable artifact categories into four temporally ordered classes, namely files, registry, process, and network (green boxes). This approach allows us to measure and quantify the execution of the malware binaries on a spectrum of observable sandbox activities. The spectrum of artifacts allows us to measure the full, partial, and failed execution of malware binaries.

**Artifact Ground Truth.** We separate the artifacts into system and network categories. System artifacts are creation, modification, or deletion of system resources such as files, registry, and process mutexes. Network artifacts are network traffic that can be observed outside the analysis environment. Based on our manual analysis of the malware families, we create unique indicators from files, registries, and process mutexes for each family to identify how far each sample executed in the sandbox. We provide a sample of these signatures in Table 7 in Appendix A.

For the network artifacts, we take every domain and IP address from the sandbox execution and manually filter them based on their usage. For example, some malware will contact benign websites like *pastebin* to retrieve configurations or commands. We are aware that *pastebin* is a benign domain, but we include it in the malicious network artifact list if the request is to download content. We manually examined all popular benign domains (dropbox, google drive, one drive, etc.) and inspected the HTTP requests to label them as maliciously abused. This step is necessary to help us quantify what malware fully executed in the sandbox. Blindly removing all benign domains will underestimate the execution and, therefore, be incomplete.

We observed 6,610 domains and IP addresses across both sandboxes and after removing benign domains and requests, we were left with 5,304 domains and IP addresses.

**Quantifying Malware Execution.** Using the system signatures and network artifacts, we quantify the execution of each binary in both sandboxes. We count the number of signature matches per artifact category from the sandbox reports. The search uses disjunctive matching and chains all signatures together per category. For example, given all file signatures, our search query chains the set of signatures with an $OR$ statement. If *any* of the signatures match, we count that as an observable artifact from the sandbox execution per category. More concretely, if a malware family is known to create a mutex with a specific pattern, and we observe that pattern in the sandbox for an instance of that family, then we count the execution successful for the process category. For network category, we look for exact matches for domains and IP addresses. Since the ground truth is a manually curated list of malicious domains and IP addresses, we use exact matches.

**Identifying Malware Families.** We rely on prior works that use machine learning to classify malware families [30], [33], [38], [114]. We use similar features found in malware sandbox behavior reports to create a feature vector for each malware and evaluate the family classification accuracy, precision, recall, and F1 score. Table 2 summarizes the set of features used for the classification of malware families. The high-fidelity malware labels are given by the ground-truth and collection process outlined earlier. We evaluated several algorithms, including different configurations of SVM, logistic regression, classification trees, KNN, and perceptron. In total, we evaluated 13 different configurations of machine learning algorithms. Due to space limitations, we only present the top three classifiers; however, complete results can be found in Table 8 in Appendix B.

TABLE 2: A summary of features used by classifiers.

| Category | Features |
|---|---|
| File | Count of created, modified, deleted, file size, extensions, and paths |
| Registry | Count of created keys, modified keys, deleted keys, and key types |
| Process | Count of mutex |
| Network | Count of dst IP, ports. TCP, UDP, RAW, POST, GET, HEAD, response codes, request and response size, and DNS record types (MX, NS, A, etc.) |

### 5.5. Results

**Experiment Setup.** Our guided sandbox uses VMRay, a commercial cloud-based malware sandbox. VMRay provides a virtualized implementation and outside guest monitoring capabilities (Intel EPT hooks). The analysis environment is configurable and provides rich features that allow us to implement the guidelines mentioned earlier. Our unguided sandbox uses a default installation of the Cuckoo sandbox [98]. Cuckoo is a popular open source malware analysis sandbox that is widely used in academia and industry. Cuckoo is a good candidate for an unguided sandbox because new users may initially be looking for a free solution to experiment with. We deployed Cuckoo on a local server with 12 CPU cores, 64 GB of RAM, and 1 TB of disk storage. The analysis server is hosted on an isolated network that has internet access. Each malware sample is run for a maximum of two minutes to mitigate any potential attacks from the infected sandbox environment to other systems on the internet. We use Scikit-learn [115] to implement the classification of the malware family. Our experiment code is written in Python and consists of 545 lines of code for report parsing, feature extraction, and model training and evaluation.

TABLE 3: Summary of observed artifacts by categories in guided (G; Green) and unguided (U; Red) sandbox configurations.

| Malware Family | Sample Count | File G | File U | Registry G | Registry U | Process G | Process U | Network G | Network U |
|---|---|---|---|---|---|---|---|---|---|
| Agenttesla | 200 | ◐ | ○ | ◐ | ○ | -- | -- | ◐ | ○ |
| Amadey | 178 | ◐ | ● | ◐ | ○ | ◐ | ○ | ◐ | ○ |
| AZORult | 195 | ◐ | ◐ | ◐ | ○ | ○ | ○ | ◐ | ◐ |
| Blacknet | 200 | ◐ | ◐ | ◐ | ○ | ○ | ○ | ◐ | ○ |
| Formbook | 200 | ◐ | ◐ | ○ | ○ | ○ | ○ | ◐ | ◐ |
| Lokibot | 200 | ● | ● | ● | ○ | ● | ◐ | ● | ◐ |
| Neutrino | 98 | ◐ | ◐ | ○ | ○ | ◐ | ◐ | ◐ | ● |
| Ponystealer | 200 | ◐ | ● | ● | ● | -- | -- | ◐ | ◐ |

**Objective 1: Malware Blocklist.** To create a blocklist of malicious domains and IP addresses, the malware must run in full according to the model in Figure 2. This task only requires network artifacts. For a given family sample size (i.e. 200 binaries), we quantify how many of those generated a malicious network artifact. We refer to this number as the activity coverage for a given artifact category. Table 3 summarizes the observed network artifacts for both sandboxes (G and U) under the "Network" column. We find that the network artifact coverage for the guided sandbox surpasses that of the unguided sandbox. More precisely, we find that the guided sandbox provides, on average, $4.5x$ activity coverage compared to the unguided sandbox. For example, 47.5% of the samples for *Agenttesla* generate a malicious network artifact in the guided sandbox versus 0% for the unguided.

However, the coverage within the guided sandbox varies by malware family. For example, *Amadey* and *Lokibot* have 95.51% and 94% coverage, respectively. On the other hand, *Blacknet* and *Formbook* have 28% and 26.5% coverage, respectively. A closer inspection of *Formbook* revealed that the malware uses a set of aggressive anti-analysis tactics like delayed execution, detect user-space and kernel-space debuggers, detect virtual machines, and evade system call hooks. We observe that 70.5% of *Formbook* fail to execute due to anti-analysis. Approximately 17% of those samples detect the virtualization environment and 97% try to detect the monitoring tools.

For *Blacknet*, we find that malware crashes due to missing system dependencies and incompatibilities with Windows 10. Specifically, 72% of *Blacknet* samples crash on execution. In addition, 50% of the samples produced empty system call logs, suggesting that the executable file was not loaded into memory. Finally, we find that the guided sandbox provides coverage for most malware families but has noticeably less coverage for newer malware families such as *Agenttesla* (47.5%) and *Formbook* (26.5%). This is also apparent in the unguided sandbox, where the older malware families, such as *Neutrion* (37.76%) and *Ponystealer* (35.5%), have more coverage than the more recently active families.

**Objective 2: Malware Behavior.** Extracting malware behavior is the most common objective according to Yong Wong et al. [25]. Malware behavior involves both system and network artifacts. We leverage our hand-crafted signatures based on manual analysis to quantify the execution of malware families. By inspecting all four categories independently, we can measure the observability of malware behavior. Table 3 presents the coverage observed per category for both sandboxes. We find a significant difference between guided and unguided sandboxes. First, the file category shows that the guided sandbox provides $1.6x$ better behavior coverage than the unguided sandbox. In particular, the biggest difference can be seen for *Agenttesla*. We observe *Agenttesla* file artifacts in 74.5% of the sample set for the guided sandbox compared to only one sample (0.5%) seen in the unguided configuration. For *Neutrino*, the unguided sandbox appears to have more file activity, but upon closer inspection, we discovered that these activities are associated with anti-analysis behavior.

Second, the registry activity of the guided sandbox shows, on average, $4.6x$ better coverage than that of the unguided sandbox. The unguided sandbox only had matches for *Neutrino* and *Ponystealer*. We find that *Ponystealer* is the only family that has a similar coverage of registry activity in both guided (76.5%) and unguided (71%) sandboxes. *Ponystealer* creates similar registry artifacts, but fails to execute network communication because the malware detects the virtual machine environment (anti-VM) in the unguided setup. The remaining families had no observable activities in the unguided results. Third, the process activity category only considers mutex creation. *Agenttesla* and *Ponystealer* do not create mutex artifacts; therefore, we could not quantify them. For the other families, we observe that the guided sandbox provides $11.3x$ more process mutex activity than the unguided sandbox. The unguided sandbox reports show mutex activity for only *Lokibot* (10.67%) and *Neutrino* (14.29%). On the other hand, the guided sandbox had $9x$ and $1.7x$ more process activity (mutex) for *Lokibot* (94.67%) and *Neutrino* (24.49%), respectively.

Our manual analysis of the malware families with source code shows that mutex is used to ensure that only one instance of the malware is running at a time to avoid multiple registration with the C&C. However, not all instances of malware rely on this feature, and some reports that have network activities (full execution) do not have mutex activities, even though it is expected. We believe that this can be attributed to custom variants of the malware family since some families have source code available. Nevertheless, we can see a significant difference, empirically, between the guided and unguided sandbox. When comparing the execution flow across families for the guided sandbox, we observe variations in activities. For example, 47.5% of *Agenttesla* samples attempt to communicate with their C&C while 74% of the samples exhibit file and registry activities. This indicates a partial execution for the remaining 26% of samples.

A closer look shows that some *Agenttesla* samples exhibit slightly different behavior due to the configurations given by the malware operator. Similarly, *Amadey* can profile the infected system for AV software; however, we only observe this behavior in some samples, since this would require the malware operator to activate the feature. These customizations can create additional artifacts that our signatures do not capture. This is especially the case for malware samples that are configured to download additional malware (first-stage downloaders). We would observe the initial persistence stage, then a single capability that downloads and executes additional malware.

TABLE 4: A summary of the classification results for the top three best preforming algorithms when using guided, unguided, and combining sandbox artifacts.

| Sandbox Type | Algorithm | A | P | R | F1 |
|---|---|---|---|---|---|
| Guided | Random Forest | 96.52 | 96.77 | 95.30 | 96.57 |
| | Decision Trees | 92.17 | 92.38 | 89.40 | 92.18 |
| | KNN | 90.43 | 90.66 | 88.00 | 90.41 |
| Unguided | Random Forest | 70.31 | 69.21 | 62.80 | 68.43 |
| | Decision Trees | 67.69 | 67.03 | 62.00 | 67.01 |
| | KNN | 65.07 | 65.47 | 60.20 | 64.74 |
| Combined | Random Forest | 97.40 | 97.50 | 96.10 | 97.40 |
| | Decision Trees | 94.81 | 94.96 | 91.90 | 94.80 |
| | KNN | 92.86 | 93.06 | 90.80 | 92.87 |
| | Improvement | 0.88 | 0.73 | 0.78 | 0.83 |

**Objective 3: Malware Families.** We apply a machine learning approach to classify malware families based on sandbox artifacts. Table 4 presents the classification results (in percentages) for the three top algorithms that perform the best. We find a significant difference between the guided and unguided sandbox results in terms of accuracy (A), precision (P), recall (R), and F1 score. The random forest algorithm performed the best out of the 13 different configurations. We provide the complete results in Table 8 in Appendix B. When we combine the artifacts of both sandboxes, we get a slight improvement of less than 1%. We expect the guided artifacts to outperform the unguided artifacts simply because the signal to differentiate between malware families is present. Using the combine features of both sandboxes, we see a negligible improvement.

To understand why this is the case, we performed a Principal Component Analysis (PCA) to understand the importance of the classification features. The five most important features of the guided sandbox include file creation, network response size, file extension, registry creation, and HTTP method. The five most important features of the unguided sandbox include file size, mutex, file path, HTTP method, and type of registry keys. Cross referencing those features with Table 3 gives us more context about the classification. For the guided, we find that the file, registry, and network categories provide complementary activity coverage for the malware families. For example, when a malware family (*Formbook*) lacks one or more activity categories (file and registry), the other categories (process and network) complement them and help classify them.

For the unguided, the file, network, and process categories rank higher in importance for classification, even though the process artifacts (mutex) have low coverage. Furthermore, the lack of activity coverage for some malware families in the unguided sandbox could be contributing to the classifier accuracy of 70%. In other words, the classifier correctly classifies malware families with 70% accuracy based on the wrong features. The phenomenon is known as the base rate fallacy [52] and underscores the impact an unguided sandbox configuration can have on security applications such as malware family classification. The absence of activity and its effect on the classifier's performance (and perceived findings) is even more evident in Table 8 in Appendix B.

## 6. Discussion & Conclusion

To recap, the guided sandbox applied as much of the proposed guidelines as possible for the use cases we studied. Using this experiment setup, we quantified the activity coverage for creating a blocklist, extracting malware behavior, and labeling malware families.

The blocklist objective shows that even with a well-configured sandbox, we can expect failed execution, anti-analysis, and missing dependencies. Moreover, the age of the malware can influence the blocklist results. For example, we observed that older malware like *Neutrino* had more activity coverage than newer malware like *Formbook*, primarily because newer malware tend to utilize more effective tactics that can evade sandbox analysis.

The malware behavior objective shows that malware execution is non-deterministic in nature. This non-determinism can be partially attributed to the malware configuration and customization by the operator. For example, we observed several malware samples from the same family (*Agenttesla* and *Amadey*) to have varied behavior. These variations are noticeable in the observed file and registry artifacts. One report of *Agenttesla* mentions that that particular sample is configured to search and steal crytpocurrency wallets, while another sample did not exhibit that behavior. Thus, the usage of multiple artifact categories (file, registry, process, and network) can provide a better understanding of how malware executes.

Finally, the malware family objective shows that classifiers can perform poorly when they lack observable artifacts. Our empirical experiment shows that the guided sandbox artifacts provide much higher fidelity classification results in comparison with the unguided. We also show that combining artifacts from two or more sandboxes can potentially improve the classification results. We conjecture a better configured sandbox ensemble can provide additional artifacts for classification because even with the unguided sandbox we observed some improvement. Finally, our experiments shed light on reasons why family classifiers can achieve a higher than expected accuracy based on latent signals or base rate fallacy [116]. For example, classifiers can incorrectly incorporate features based on the lack of artifacts observed for a given family. Therefore, the classifier may achieve a higher than expected accuracy, since the classification algorithms learn the family label based on the lack of observed artifacts.

Taken as a whole, our evaluations affirm that there is no "silver bullet" for configuring and using sandboxes. To ensure that end-users use sandboxes in ways that provide the best chance of collecting meaningful artifacts, we recommend that they define an appropriate scope for their analyses, a corresponding threat model, and derive context about how artifacts collected from the sandbox could influence their intended use case. Overall, as in-depth assessments show that care must be taken when using sandboxes for security applications, keeping in mind that even when applying advanced techniques, using multiple analyses, and systematically testing their configurations, the results can still be incomplete due to limitations of the underlying technologies. For these reasons, it is imperative that end-users document these limitations and identify possible remedies based on the guidelines we suggested.

# References

[1] L. Bilge, E. Kirda, C. Kruegel, and M. Balduzzi, "Exposure: Finding malicious domains using passive dns analysis.," in *Proc. of the 18th NDSS*, San Diego, CA, Feb. 2011.

[2] Y. Sun, K. Jee, S. Sivakorn, et al., "Detecting malware injection with program-dns behavior," in *Proc. of the 5th EuroS&P*, Sep. 2020.

[3] D. Balzarotti, M. Cova, C. Karlberger, E. Kirda, C. Kruegel, and G. Vigna, "Efficient detection of split personalities in malware.," in *Proc. of the 17th NDSS*, San Diego, CA, Feb. 2010.

[4] G. Jacob, R. Hund, C. Kruegel, and T. Holz, "Jackstraws: Picking command and control connections from bot traffic.," in *Proc. of the 20th USENIX Security*, San Francisco, CA, Aug. 2011.

[5] E. Cozzi, M. Graziano, Y. Fratantonio, and D. Balzarotti, "Understanding linux malware," in *Proc. of the 39th S&P Oakland*, San Francisco, CA, May 2018.

[6] A. R. A. Grégio, P. L. De Geus, C. Kruegel, and G. Vigna, "Tracking memory writes for malware classification and code reuse identification," in *Proc. of the DIMVA*, Jul. 2012.

[7] J. Caballero, P. Poosankam, C. Kreibich, and D. Song, "Dispatcher: Enabling active botnet infiltration using automatic protocol reverse-engineering," in *Proc. of the 16th ACM CCS*, Chicago, Illinois, Nov. 2009.

[8] X. Ugarte-Pedrero, D. Balzarotti, I. Santos, and P. G. Bringas, "Sok: Deep packer inspection: A longitudinal study of the complexity of run-time packers," in *Proc. of the 36th S&P Oakland*, San Jose, CA, May 2015.

[9] D. Kirat and G. Vigna, "Malgene: Automatic extraction of malware analysis evasion signature," in *Proc. of the 22nd ACM CCS*, Denver, Colorado, Oct. 2015.

[10] F. Barr-Smith, X. Ugarte-Pedrero, M. Graziano, R. Spolaor, and I. Martinovic, "Survivalism: Systematic analysis of windows malware living-off-the-land," in *Proc. of the 42nd S&P Oakland*, Online Conference, May 2021.

[11] M. Egele, T. Scholte, E. Kirda, and C. Kruegel, "A survey on automated dynamic malware-analysis techniques and tools," *ACM computing surveys (CSUR)*, 2012.

[12] O. Or-Meir, N. Nissim, Y. Elovici, and L. Rokach, "Dynamic malware analysis in the modern era—a state of the art survey," *ACM Computing Surveys (CSUR)*, 2019.

[13] C. Rossow, C. J. Dietrich, C. Grier, et al., "Prudent practices for designing malware experiments: Status quo and outlook," in *Proc. of the 33rd S&P Oakland*, San Francisco, CA, May 2012.

[14] A. Bulazel and B. Yener, "A survey on automated dynamic malware analysis evasion and counter-evasion: Pc, mobile, and web," in *Proceedings of the 1st ACM Reversing and Offensive-oriented Trends Symposium*, 2017.

[15] D. C. D'Elia, E. Coppa, S. Nicchi, F. Palmaro, and L. Cavallaro, "Sok: Using dynamic binary instrumentation for security (and how you may get caught red handed)," in *Proc. of the 14th ACM Symposium on Information, Computer and Communications Security (ASIACCS)*, Auckland, New Zealand, Apr. 2019.

[16] D. C. D'Elia, E. Coppa, F. Palmaro, and L. Cavallaro, "On the dissection of evasive malware," *IEEE Transactions on Information Forensics and Security*, 2020.

[17] M. Egele, C. Kruegel, E. Kirda, H. Yin, and D. Song, "Dynamic spyware analysis," in *Proc. of the 2007 USENIX Annual Technical Conference (ATC)*, Santa Clara, CA, Jun. 2007.

[18] QEMU, *QEMU: the FAST! processor emulator*, https://www.qemu.org, 2019.

[19] Xen Project, *WHAT IS THE XEN PROJECT?* https://xenproject.org/about-us/, Online; accessed 25 January 2020.

[20] M. Jones, *Virtio: An I/O virtualization framework for Linux*, https://developer.ibm.com/articles/l-virtio/, Jan. 2010.

[21] B. Jain, M. B. Baig, D. Zhang, D. E. Porter, and R. Sion, "Sok: Introspections on trust and the semantic gap," in *Proc. of the 35th S&P Oakland*, San Jose, CA, May 2014.

[22] M. N. Hossain, S. M. Milajerdi, J. Wang, et al., "$SLEUTH$: Real-time attack scenario reconstruction from $COTS$ audit data," in *Proc. of the 26th USENIX Security*, Vancouver, BC, Canada, Aug. 2017.

[23] N. Miramirkhani, M. P. Appini, N. Nikiforakis, and M. Polychronakis, "Spotless sandboxes: Evading malware analysis systems using wear-and-tear artifacts," in *Proc. of the 38th S&P Oakland*, San Jose, CA, May 2017.

[24] C. Xuan, J. Copeland, and R. Beyah, "Toward revealing kernel malware behavior in virtual execution environments," in *Proc. of the 12th RAID*, Saint-Malo, France, Sep. 2009.

[25] M. Yong Wong, M. Landen, M. Antonakakis, D. M. Blough, E. M. Redmiles, and M. Ahamad, "An inside look into the practice of malware analysis," in *Proc. of the 28th ACM CCS*, Seoul, South Korea, Nov. 2021.

[26] D. Kim, A. Majlesi-Kupaei, J. Roy, et al., "Dynodet: Detecting dynamic obfuscation in malware," in *Proc. of the DIMVA*, Bonn, DE, Jul. 2017.

[27] E. Kirda, C. Kruegel, G. Banks, G. Vigna, and R. Kemmerer, "Behavior-based spyware detection.," in *Proc. of the 15th USENIX Security*, Vancouver, Canada, Jul. 2006.

[28] M. Antonakakis, R. Perdisci, D. Dagon, W. Lee, and N. Feamster, "Building a dynamic reputation system for dns.," in *Proc. of the 19th USENIX Security*, Washington, DC, Aug. 2010.

[29] L. Litty, H. A. Lagar-Cavilla, and D. Lie, "Hypervisor support for identifying covertly executing binaries.," in *Proc. of the 17th USENIX Security*, San Jose, CA, Aug. 2008.

[30] K. Rieck, T. Holz, C. Willems, P. Düssel, and P. Laskov, "Learning and classification of malware behavior," in *Proc. of the DIMVA*, Jul. 2008.

[31] W.-J. Li, S. Stolfo, A. Stavrou, E. Androulaki, and A. D. Keromytis, "A study of malcode-bearing documents," in *Proc. of the DIMVA*, Lucerne, CH, Jul. 2007.

[32] W. Cui, M. Peinado, Z. Xu, and E. Chan, "Tracking rootkit footprints with a practical memory analysis system," in *Proc. of the 21st USENIX Security*, Bellevue, WA, Aug. 2012.

[33] U. Bayer, P. M. Comparetti, C. Hlauschek, C. Kruegel, and E. Kirda, "Scalable, behavior-based malware clustering," in *Proc. of the 16th NDSS*, San Diego, CA, Feb. 2009.

[34] M. Z. Rafique and J. Caballero, "Firma: Malware clustering and network signature generation with mixed network behaviors," in *Proc. of the 16th RAID*, St. Lucia, Sep. 2013.

[35] F. Tegeler, X. Fu, G. Vigna, and C. Kruegel, "Botfinder: Finding bots in network traffic without deep packet inspection," in *Proceedings of the 8th international conference on Emerging networking experiments and technologies*, 2012.

[36] R. Perdisci, W. Lee, and N. Feamster, "Behavioral clustering of http-based malware and signature generation using malicious network traces.," in *Proc. of the 7th USENIX Symposium on Networked Systems Design and Implementation (NSDI)*, San Jose, CA, Apr. 2010.

[37] T.-F. Yen and M. K. Reiter, "Traffic aggregation for malware detection," in *Proc. of the DIMVA*, Jul. 2008.

[38] M. Bailey, J. Oberheide, J. Andersen, Z. M. Mao, F. Jahanian, and J. Nazario, "Automated classification and analysis of internet malware," in *International Workshop on Recent Advances in Intrusion Detection*, Queensland, Australia, Sep. 2007.

[39] C. Kolbitsch, T. Holz, C. Kruegel, and E. Kirda, "Inspector gadget: Automated extraction of proprietary gadgets from malware binaries," in *Proc. of the 31th S&P Oakland*, Oakland, CA, May 2010.

[40] R. Paleari, L. Martignoni, E. Passerini, et al., "Automatic generation of remediation procedures for malware infections.," in *Proc. of the 19th USENIX Security*, Washington, DC, Aug. 2010.

[41] X. Hu and K. G. Shin, "Duet: Integration of dynamic and static analyses for malware clustering with cluster ensembles," in *Proc. of the 29th ACSAC*, 2013.

[42] M. Lindorfer, C. Kolbitsch, and P. M. Comparetti, "Detecting environment-sensitive malware," in *Proc. of the 14th RAID*, Menlo Park, California, Sep. 2011.

[43] P. Vadrevu and R. Perdisci, "Maxs: Scaling malware execution with sequential multi-hypothesis testing," in *Proc. of the 11th ACM Symposium on Information, Computer and Communications Security (ASIACCS)*, Xi'an, China, Jun. 2016.

[44] U. Bayer, E. Kirda, and C. Kruegel, "Improving the efficiency of dynamic malware analysis," in *Proc. of the 2010 ACM Symposium on Applied Computing (SAC)*, Sierre, Switzerland, Mar. 2010.

[45] D. Kirat and G. Vigna, "Barecloud: Bare-metal analysis-based evasive malware detection," in *Proc. of the 23rd USENIX Security*, San Diego, CA, Aug. 2014.

[46] T. Raffetseder, C. Kruegel, and E. Kirda, "Detecting system emulators," in *International Conference on Information Security*, Springer, 2007.

[47] X. Chen, J. Andersen, Z. M. Mao, M. Bailey, and J. Nazario, "Towards an understanding of anti-virtualization and anti-debugging behavior in modern malware," in *Proc. of the International Conference on Dependable Systems and Networks (DSN)*, 2008.

[48] K. A. Roundy and B. P. Miller, "Hybrid analysis and control of malware," in *Proc. of the 13th RAID*, Ottawa, Canada, Sep. 2010.

[49] C. Willems, R. Hund, A. Fobian, D. Felsch, T. Holz, and A. Vasudevan, "Down to the bare metal: Using processor features for binary analysis," in *Proc. of the 28th ACSAC*, 2012.

[50] Y. Kawakoya, M. Iwamura, E. Shioji, and T. Hariu, "Api chaser: Anti-analysis resistant malware analyzer," in *Proc. of the 16th RAID*, St. Lucia, Sep. 2013.

[51] Z. Xu, J. Zhang, G. Gu, and Z. Lin, "Goldeneye: Efficiently and effectively unveiling malware's targeted environment," in *Proc. of the 17th RAID*, Gothenburg, Sweden, Sep. 2014.

[52] A. Küchler, A. Mantovani, Y. Han, L. Bilge, and D. Balzarotti, "Does every second count? time-based evolution of malware behavior in sandboxes," in *Proc. of the 2021 NDSS*, Virtual, Feb. 2021.

[53] P. Pawlak and C. Wendling, "Trends in cyberspace: Can governments keep up?" *Environment Systems and Decisions*, 2013.

[54] T. Grant, "On the military geography of cyberspace," *Leading Issues in Cyber Warfare and Security: Cyber Warfare Secur*, 2015.

[55] P. Pawlak, "Capacity building in cyberspace as an instrument of foreign policy," *Global Policy*, 2016.

[56] S. Dambra, L. Bilge, and D. Balzarotti, "Sok: Cyber insurance–technical challenges and a system security roadmap," in *Proc. of the 41st S&P Oakland*, May 2020.

[57] A. Neupane, M. L. Rahman, N. Saxena, and L. Hirshfield, "A multi-modal neuro-physiological study of phishing detection and malware warnings," in *Proc. of the 22nd ACM CCS*, Denver, Colorado, Oct. 2015.

[58] M. Christodorescu, S. Jha, and C. Kruegel, "Mining specifications of malicious behavior," in *Proc. of the 6th joint meeting of European Software Engineering Conference (ESEC) and ACM SIGSOFT Symposium on the Foundations of Software Engineering (FSE)*, Dubrovnik, Croatia, Sep. 2007.

[59] L. Martignoni, E. Stinson, M. Fredrikson, S. Jha, and J. C. Mitchell, "A layered architecture for detecting malicious behaviors," in *Proc. of the 11th RAID*, Cambridge, Massachusetts, Sep. 2008.

[60] M. Fredrikson, S. Jha, M. Christodorescu, R. Sailer, and X. Yan, "Synthesizing near-optimal malware specifications from suspicious behaviors," in *Proc. of the 31th S&P Oakland*, Oakland, CA, May 2010.

[61] B. Anderson, D. Quist, J. Neil, C. Storlie, and T. Lane, "Graph-based malware detection using dynamic analysis," *Journal in computer Virology*, 2011.

[62] C. Willems, F. C. Freiling, and T. Holz, "Using memory management to detect and extract illegitimate code for malware analysis," in *Proc. of the 28th ACSAC*, 2012.

[63] S. Palahan, D. Babić, S. Chaudhuri, and D. Kifer, "Extraction of statistically significant malware behaviors," in *Proc. of the 29th ACSAC*, 2013.

[64] R. Canzanese, S. Mancoridis, and M. Kam, "Run-time classification of malicious processes using system call analysis," in *Proceedings of the 10th International Conference on Malicious and Unwanted Software (MALWARE)*, 2015.

[65] A. Kharraz, S. Arshad, C. Mulliner, W. Robertson, and E. Kirda, "$UNVEIL$: A large-scale, automated approach to detecting ransomware," in *Proc. of the 25th USENIX Security*, Austin, TX, Aug. 2016.

[66] M. Xu and T. Kim, "Platpal: Detecting malicious behavior with platform diversity," in *Proc. of the 26th USENIX Security*, Vancouver, BC, Canada, Aug. 2017.

[67] J. Zhang, Z. Gu, J. Jang, *et al.*, "Scarecrow: Deactivating evasive malware via its own evasive logic," in *Proc. of the International Conference on Dependable Systems and Networks (DSN)*, 2020.

[68] P. Royal, M. Halpin, D. Dagon, R. Edmonds, and W. Lee, "Polyunpack: Automating the hidden-code extraction of unpack-executing malware," in *Proc. of the 22nd ACSAC*, 2006.

[69] M. Abu Rajab, J. Zarfoss, F. Monrose, and A. Terzis, "A multifaceted approach to understanding the botnet phenomenon," in *Proc. of the 6th ACM SIGCOMM Conference on Internet Measurement (IMC)*, 2006.

[70] M. G. Kang, P. Poosankam, and H. Yin, "Renovo: A hidden code extractor for packed executables," in *Proceedings of the 2007 ACM workshop on Recurring malcode*, 2007.

[71] M. Polychronakis, P. Mavrommatis, and N. Provos, "Ghost turns zombie: Exploring the life cycle of web-based malware," 2008.

[72] T. Holz, M. Steiner, F. Dahl, E. Biersack, F. C. Freiling, *et al.*, "Measurements and mitigation of peer-to-peer-based botnets: A case study on storm worm.," in *Proceedings of the 2008 USENIX Conference on Large-Scale Exploits & Emergent Threats (USENIX LEET)*, 2008.

[73] M. Sharif, A. Lanzi, J. Giffin, and W. Lee, "Automatic reverse engineering of malware emulators," in *Proc. of the 30th S&P Oakland*, Oakland, CA, May 2009.

[74] B. Stone-Gross, C. Kruegel, K. Almeroth, A. Moser, and E. Kirda, "Fire: Finding rogue networks," in *Proc. of the 25th ACSAC*, 2009.

[75] D. A. Quist and L. M. Liebrock, "Visualizing compiled executables for malware analysis," in *Proceedings of the 6th IEEE International Workshop on Visualization for Cyber Security*, 2009.

[76] U. Bayer, I. Habibi, D. Balzarotti, E. Kirda, and C. Kruegel, "A view on current malware behaviors," in *Proceedings of the 2008 USENIX Conference on Large-Scale Exploits & Emergent Threats (USENIX LEET)*, 2009.

[77] T. Holz, M. Engelberth, and F. Freiling, "Learning more about the underground economy: A case-study of keyloggers and dropzones," in *Proc. of the 14th ESORICS*, Saint Malo, France, Sep. 2009.

[78] S. Krishnan, K. Z. Snow, and F. Monrose, "Trail of bytes: Efficient support for forensic analysis," in *Proc. of the 17th ACM CCS*, Chicago, Illinois, Oct. 2010.

[79] P. M. Comparetti, G. Salvaneschi, E. Kirda, C. Kolbitsch, C. Kruegel, and S. Zanero, "Identifying dormant functionality in malware programs," in *Proc. of the 31th S&P Oakland*, Oakland, CA, May 2010.

[80] C. Leita, U. Bayer, and E. Kirda, "Exploiting diverse observation perspectives to get insights on the malware landscape," in *Proc. of the International Conference on Dependable Systems and Networks (DSN)*, 2010.

[81] Y. Nadji, M. Antonakakis, R. Perdisci, and W. Lee, "Understanding the prevalence and use of alternative plans in malware with network games," in *Proc. of the 27th ACSAC*, 2011.

[82] M. Neugschwandtner, P. M. Comparetti, and C. Platzer, "Detecting malware's failover c&c strategies with squeeze," in *Proc. of the 27th ACSAC*, 2011.

[83] J. Caballero, C. Grier, C. Kreibich, and V. Paxson, "Measuring pay-per-install: The commoditization of malware distribution.," in *Proc. of the 20th USENIX Security*, San Francisco, CA, Aug. 2011.

[84] C. Rossow, C. Dietrich, and H. Bos, "Large-scale analysis of malware downloaders," in *Proc. of the DIMVA*, Jul. 2012.

[85] M. Lindorfer, A. Di Federico, F. Maggi, P. M. Comparetti, and S. Zanero, "Lines of malicious code: Insights into the malicious software industry," in *Proc. of the 28th ACSAC*, 2012.

[86] N. Jagpal, E. Dingle, J.-P. Gravel, *et al.*, "Trends and lessons from three years fighting malicious extensions," in *Proc. of the 24th USENIX Security*, Washington, DC, Aug. 2015.

[87] A. Kharraz, W. Robertson, D. Balzarotti, L. Bilge, and E. Kirda, "Cutting the gordian knot: A look under the hood of ransomware attacks," in *Proc. of the DIMVA*, Milan, IT, Jul. 2015.

[88] B. Yadegari, B. Johannesmeyer, B. Whitely, and S. Debray, "A generic approach to automatic deobfuscation of executable code," in *Proc. of the 36th S&P Oakland*, San Jose, CA, May 2015.

[89] K. Thomas, E. Bursztein, C. Grier, *et al.*, "Ad injection at scale: Assessing deceptive advertisement modifications," in *Proc. of the 36th S&P Oakland*, San Jose, CA, May 2015.

[90] X. Han, N. Kheir, and D. Balzarotti, "The role of cloud services in malicious software: Trends and insights," in *Proc. of the DIMVA*, Milan, IT, Jul. 2015.

[91] X. Ugarte-Pedrero, D. Balzarotti, I. Santos, and P. G. Bringas, "Rambo: Run-time packer analysis with multiple branch observation," in *Proc. of the DIMVA*, Donostia-San Sebastián, ES, Jul. 2016.

[92] K. Thomas, J. A. E. Crespo, R. Rasti, *et al.*, "Investigating commercial pay-per-install and the distribution of unwanted software," in *Proc. of the 25th USENIX Security*, Austin, TX, Aug. 2016.

[93] C. Lever, P. Kotzias, D. Balzarotti, J. Caballero, and M. Antonakakis, "A lustrum of malware network communication: Evolution and insights," in *Proc. of the 38th S&P Oakland*, San Jose, CA, May 2017.

[94] B. Farinholt, M. Rezaeirad, P. Pearce, *et al.*, "To catch a ratter: Monitoring the behavior of amateur darkcomet rat operators in the wild," in *Proc. of the 38th S&P Oakland*, San Jose, CA, May 2017.

[95] I. Haq, S. Chica, J. Caballero, and S. Jha, "Malware lineage in the wild," *Computers & Security*, 2018.

[96] E. Avllazagaj, Z. Zhu, L. Bilge, D. Balzarotti, and T. Dumitras, "When malware changed its mind: An empirical study of variable program behaviors in the real world," in *Proc. of the 30th USENIX Security*, Aug. 2021.

[97] A. Dinaburg, P. Royal, M. Sharif, and W. Lee, "Ether: Malware analysis via hardware virtualization extensions," in *Proc. of the 15th ACM CCS*, Alexandria, VA, Oct. 2008.

[98] C. Guarnieri, A. Tanasi, J. Bremer, and M. Schloesser, "The cuckoo sandbox," *Accessed: Dec*, 2012.

[99] U. Bayer, C. Kruegel, and E. Kirda, "TTAnalyze: A tool for analyzing malware," in *European Institute for Computer Antivirus Research (EICAR)*, 2006.

[100] C. Willems, T. Holz, and F. Freiling, "Toward automated dynamic malware analysis using cwsandbox," in *Proc. of the 28th S&P Oakland*, Oakland, CA, May 2007.

[101] A. Yokoyama, K. Ishii, R. Tanabe, *et al.*, "Sandprint: Fingerprinting malware sandboxes to provide intelligence for sandbox evasion," in *Proc. of the 19th RAID*, Evry, France, Sep. 2016.

[102] C. Rossow, C. J. Dietrich, H. Bos, *et al.*, "Sandnet: Network traffic analysis of malicious software," in *ACM Proceedings of the 1st Workshop BADGERS*, 2011.

[103] M. Bailey, D. Dittrich, E. Kenneally, and D. Maughan, "The menlo report," *IEEE Security & Privacy*, 2012.

[104] L. Gelinas, A. Wertheimer, and F. G. Miller, "When and why is research without consent permissible?" *Hastings Center Report*, 2016.

[105] M. ATT&CK, "Mitre ATT&CK," https://attack.mitre.org, 2021.

[106] *Malware Bazaar - Statistics*, https://bazaar.abuse.ch/statistics/, 2021.

[107] *CyberCrime Tracker Stats*, http://cybercrime-tracker.net/stats.php, 2021.

[108] *Malware Trends Tracker*, https://any.run/malware-trends/, 2021.

[109] *Malware bazaar - yara*. [Online]. Available: %5Curl%7Bhttps://bazaar.abuse.ch/sample/efe947e0a8842997d152af946ef0293a972cc11662f3c62a8461bc4a07427669/#yara%7D.

[110] V. DÍAZ, *Context is king (part i) - crowdsourced sigma rules*, May 2021. [Online]. Available: %5Curl%7Bhttps://blog.virustotal.com/2021/05/context-is-king-part-i-crowdsourced.html%7D.

[111] *Snort- rules*. [Online]. Available: %5Curl%7Bhttps://www.snort.org/downloads#rules%7D.

[112] M. Sebastián, R. Rivera, P. Kotzias, and J. Caballero, "Avclass: A tool for massive malware labeling," in *Proc. of the 19th RAID*, Evry, France, Sep. 2016.

[113] O. Alrawi, C. Lever, K. Valakuzhy, K. Snow, F. Monrose, M. Antonakakis, et al., "The circle of life: A $Large-Scale$ study of the $IoT$ malware lifecycle," in *Proc. of the 30th USENIX Security*, Aug. 2021.
[114] A. Mohaisen, O. Alrawi, and M. Mohaisen, "Amal: High-fidelity, behavior-based automated malware analysis and classification," *Computers & Security*, 2015.
[115] F. Pedregosa, G. Varoquaux, A. Gramfort, et al., "Scikit-learn: Machine learning in python," *the Journal of machine Learning research*, vol. 12, 2011.
[116] D. Arp, E. Quiring, F. Pendlebury, et al., "Dos and don'ts of machine learning in computer security," in *Proc. of the 31th USENIX Security*, Aug. 2022.
[117] G. W. Dunlap, S. T. King, S. Cinar, M. A. Basrai, and P. M. Chen, "Revirt: Enabling intrusion analysis through virtual-machine logging and replay," in *Proc. of the ACM SIGOPS Operating System Review*, vol. 36, Mar. 2002.
[118] X. Jiang, D. Xu, H. J. Wang, and E. H. Spafford, "Virtual playgrounds for worm behavior investigation," in *Proc. of the 8th RAID*, Seattle, Washington, Sep. 2005.
[119] A. Moser, C. Kruegel, and E. Kirda, "Exploring multiple execution paths for malware analysis," in *Proc. of the 28th S&P Oakland*, Oakland, CA, May 2007.
[120] X. Jiang, X. Wang, and D. Xu, "Stealthy malware detection through vmm-based "out-of-the-box" semantic view reconstruction," in *Proc. of the 14th ACM CCS*, Alexandria, VA, Nov. 2007.
[121] H. Yin, D. Song, M. Egele, C. Kruegel, and E. Kirda, "Panorama: Capturing system-wide information flow for malware detection and analysis," in *Proc. of the 14th ACM CCS*, Alexandria, VA, Nov. 2007.
[122] J. Van Randwyk, K. Chiang, L. Lloyd, and K. Vanderveen, "Farm: An automated malware analysis environment," in *IEEE Proceedings of 42nd Annual International Carnahan Conference on Security Technology*, 2008.
[123] A. Lanzi, M. I. Sharif, and W. Lee, "K-tracer: A system for extracting kernel malware behavior.," in *Proc. of the 16th NDSS*, San Diego, CA, Feb. 2009.
[124] A. M. Nguyen, N. Schear, H. Jung, A. Godiyal, S. T. King, and H. D. Nguyen, "Mavmm: Lightweight and purpose built vmm for malware analysis," in *Proc. of the 25th ACSAC*, 2009.
[125] J. P. John, A. Moshchuk, S. D. Gribble, A. Krishnamurthy, et al., "Studying spamming botnets using botlab.," in *Proc. of the 6th USENIX Symposium on Networked Systems Design and Implementation (NSDI)*, Boston, MA, Apr. 2009.
[126] R. Riley, X. Jiang, and D. Xu, "Multi-aspect profiling of kernel rootkit behavior," in *Proceedings of the 4th ACM European conference on Computer systems*, 2009.
[127] B. M. Bowen, P. Prabhu, V. P. Kemerlis, S. Sidiroglou, A. D. Keromytis, and S. J. Stolfo, "Botswindler: Tamper resistant injection of believable decoys in vm-based hosts for crimeware detection," in *Proc. of the 13th RAID*, Ottawa, Canada, Sep. 2010.
[128] M. Neugschwandtner, C. Platzer, P. Milani Comparetti, and U. Bayer, "Danubis - dynamic device driver analysis based on virtual machine introspection," in *Proc. of the DIMVA*, Jul. 2010.
[129] H. Yin and D. Song, "TEMU: Binary code analysis via whole-system layered annotative execution," Electrical Engineering and Computer Sciences University of California at Berkeley, Tech. Rep., 2010.
[130] C. Kolbitsch, E. Kirda, and C. Kruegel, "The power of procrastination: Detection and mitigation of execution-stalling malicious code," in *Proc. of the 18th ACM CCS*, Chicago, Illinois, Oct. 2011.
[131] D. Kirat, G. Vigna, and C. Kruegel, "Barebox: Efficient malware analysis on bare-metal," in *Proc. of the 27th ACSAC*, 2011.
[132] C. Kreibich, N. Weaver, C. Kanich, W. Cui, and V. Paxson, "Gq: Practical containment for measuring modern malware systems," in *Proc. of the 11th ACM SIGCOMM Conference on Internet Measurement (IMC)*, 2011.
[133] P. Royal, "Entrapment: Tricking malware with transparent, scalable malware analysis," in *Black Hat USA Briefings (Black Hat USA)*, Las Vegas, NV, Aug. 2012.
[134] L.-K. Yan, M. Jayachandra, M. Zhang, and H. Yin, "V2e: Combining hardware virtualization and software emulation for transparent and extensible malware analysis," in *Proceedings of the 8th ACM SIGPLAN/SIGOPS Conference on Virtual Execution Environments*, 2012.
[135] Z. Deng, X. Zhang, and D. Xu, "Spider: Stealthy binary program instrumentation and debugging via hardware virtualization," in *Proc. of the 29th ACSAC*, 2013.
[136] A. Henderson, A. Prakash, L. K. Yan, et al., "Make it work, make it right, make it fast: Building a platform-neutral whole-system dynamic binary analysis platform," in *Proc. of the International Symposium on Software Testing and Analysis (ISSTA)*, San Jose, CA, Jul. 2014.
[137] T. K. Lengyel, S. Maresca, B. D. Payne, G. D. Webster, S. Vogl, and A. Kiayias, "Scalability, fidelity and stealth in the drakvuf dynamic malware analysis system," in *Proc. of the 30th ACSAC*, 2014.
[138] C. Spensky, H. Hu, and K. Leach, "LO-PHI: Low-observable physical host instrumentation for malware analysis.," in *Proc. of the 2016 NDSS*, San Diego, CA, Feb. 2016.
[139] D. Korczynski and H. Yin, "Capturing malware propagations with code injections and code-reuse attacks," in *Proc. of the 24th ACM CCS*, Dallas, TX, Oct. 2017.
[140] G. Severi, T. Leek, and B. Dolan-Gavitt, "Malrec: Compact full-trace malware recording for retrospective deep analysis," in *Proc. of the DIMVA*, Paris, FR, Jul. 2018.
[141] M. N. Arefi, G. Alexander, H. Rokham, et al., "Faros: Illuminating in-memory injection attacks via provenance-based whole-system dynamic information flow tracking," in *Proc. of the International Conference on Dependable Systems and Networks (DSN)*, 2018.
[142] A. Davanian, Z. Qi, and Y. Qu, "Decaf++: Elastic whole-system dynamic taint analysis," in *Proc. of the 22nd RAID*, Beijing, China, Sep. 2018.
[143] R. Shipp, *Online Scanners and Sandboxes*, https://github.com/rshipp/awesome-malware-analysis#online-scanners-and-sandboxes, Online; accessed 25 January 2020.

Malware sandbox systems describe the design and implementation of malware sandbox technology. We project the sandbox survey using Figure 1 to highlight the implementation, monitoring technique, and availability. More complete surveys on dynamic analysis (including sandboxes) can be found in prior works [11], [12], [14].

## 1. Malware Sandbox Considerations

Sandbox researchers must consider trade-offs between transparency, scalability, extensibility, and isolation. **Transparency** indicates if the analysis environment (hardware, OS, and environment) is indistinguishable from the malware's target environment. **Scalability** describes how efficiently a malware sandbox utilizes resources so that it can scale to concurrent analysis instances. **Extensibility** describes the required effort to extend a malware sandbox monitoring capability. **Isolation** describes the malware sandbox system and network isolation of an analysis environment. Each consideration plays a role in the malware analysis process, but they cannot all be satisfied.

For example, from Figure 1 a bare-metal sandbox has high hardware transparency, no isolation, expensive to scale, and is difficult to extend. On the other hand, an emulated sandbox has low hardware transparency, good isolation, moderate to scale with some overhead, and is easiest to extend. A virtualized sandbox provides moderate hardware transparency, isolation, highly scalable, and extensible. These properties provide context around malware sandbox systems that highlight their historical development. However, an important point to keep in mind is that the properties do not necessarily hold in practice due to other factors (monitoring and environment). For example, bare-metal systems provide high hardware transparency but Yokoyama et al. [101] shows that they can be identified with simple profiling.

## 2. A Survey of Malware Sandboxes

Table 5 presents a chronologically ordered list of **malware sandbox systems** found in the academic literature. We divide Table 5 into four sections, namely system information (reference and system name), implementation, monitoring, and available access (Avail.). There are many open-source and commercial malware sandboxes that are not in the academic literature. We do not include these malware sandboxes because their design is not documented. Interested readers can refer to a community-curated list of malware sandboxes [143].

TABLE 5: A summary of sandbox system papers. The available accesses (Avail.) are open-source (✓), commercial (✓$), or broken resource (✓*).

| Paper | System Name | Implementation | | | Monitoring | | | | Avail. |
|---|---|---|---|---|---|---|---|---|---|
| | | Emu. | Virt. | Metal | IGU | IGK | OGOn | OGOff | |
| Dunla02 [117] | ReVirt | | ✓ | | ✓ | | ✓ | ✓ | |
| Jiang05 [118] | vGrounds | | ✓ | | ✓ | | | | |
| Bayer06 [99] | TTAanlyze | ✓ | | | ✓ | ✓ | | | ✓$ |
| Moser07 [119] | | ✓ | | | ✓ | ✓ | ✓ | | ✓$ |
| Wille07 [100] | CWSandbox | | ✓ | | ✓ | | | | ✓$ |
| Jiang07 [120] | VMwatcher | ✓ | ✓ | | | | ✓ | | |
| Yin07 [121] | Panorama | ✓ | | | | | ✓ | ✓ | ✓ |
| Dinab08 [97] | Ether | | ✓ | | | | ✓ | | ✓* |
| Randw08 [122] | ISLAND | | | ✓ | | | ✓ | ✓ | ✓* |
| Lanzi09 [123] | K-Tracer | ✓ | | | ✓ | | | | |
| Xuan09 [24] | Rkprofiler | ✓ | | | | | ✓ | | |
| Nguye09 [124] | MAVMM | | ✓ | | ✓ | | ✓ | | |
| John09 [125] | Botlab | | ✓ | ✓ | | | | ✓ | ✓* |
| Riley09 [126] | PoKeR | | | ✓ | | | ✓ | ✓ | |
| Bowen10 [127] | BotSwindler | ✓ | | | | | ✓ | | |
| Neugs10 [128] | dAnubis | ✓ | | | | | ✓ | | ✓$ |
| Yin10 [129] | TEMU | ✓ | | | | | ✓ | ✓ | ✓ |
| Kolbi11 [130] | HASTEN | | ✓ | | ✓ | ✓ | ✓ | | ✓$ |
| Kirat11 [131] | BareBox | | | ✓ | | | ✓ | | ✓$ |
| Rosso11 [102] | Sandnet | | ✓ | | | | ✓ | ✓ | |
| Kreib11 [132] | GQ | | ✓ | ✓ | | | ✓ | | |
| Royal12 [133] | NVMTrace | | | ✓ | | | ✓ | ✓ | ✓ |
| Yan12 [134] | V2E | ✓ | ✓ | | | | ✓ | ✓ | |
| Deng13 [135] | SPIDER | | ✓ | | | | ✓ | | |
| Hende14 [136] | DECAF | ✓ | | | | | ✓ | | ✓ |
| Lengy14 [137] | DRAKVUF | | ✓ | | | | ✓ | ✓ | ✓ |
| Mohai15 [114] | AutoMal | | ✓ | | ✓ | | ✓ | | |
| Spens16 [138] | LO-PHI | ✓ | ✓ | ✓ | | | ✓ | | ✓ |
| Korcz17 [139] | Tartarus | ✓ | | | | | ✓ | ✓ | |
| Sever18 [140] | Malrec | ✓ | | | | | ✓ | ✓ | ✓* |
| Arefi18 [141] | FAROS | ✓ | | | | | ✓ | ✓ | |
| Davan19 [142] | DECAF++ | ✓ | | | | | ✓ | ✓ | ✓ |

**Implementation.** The first requirement for malware sandbox systems is to study malware in an isolated environment. Jiang et al. [118] use virtual machines to create a playground where worms can be observed and studied. Bayer et al. [99] use whole-system emulation to study malware threats. For bare-metal sandboxes, isolation can be implemented only on the network side. John et al. [125] propose a network isolation approach to ensure that malware does not infect systems outside the analysis environment. Many other systems [24], [97], [100], [120], [121], [123], [124] rely on either whole-system emulation, virtualization, or bare-metal. Each of these technologies has limitations in terms of scalability.

Sandboxes have an inevitable time cost because the malware must run for a preset time. Also, any setup time incurred based on the implementation of the sandbox adds to the overall analysis time. For instance, bare-metal-based systems [122], [125], [131]–[133], [138] incur more time in setup than emulation or virtualization. Royal et al. [133] optimize the setup process for bare-metal-based systems by leveraging ATA-over-Ethernet with copy-on-write block device mounts. Furthermore, the monitoring tools can incur additional slow-down depending on the monitoring capability (fine-grained or coarse-grained). Scaling malware sandboxes is difficult and malware sandbox designers must be mindful of bottlenecks that affect compute resources such as slow disk, low network bandwidth, and saturated memory.

**Monitoring.** Earlier malware sandboxes [99], [100], [118] rely on inside-guest user-space monitoring to capture malware behavior. Even popular open-source sandboxes, such as Cuckoo [98], rely on inside-guest user-space monitoring. The inside-guest user-space monitoring technique provides instruction-level traces but also various levels of abstraction such as system call hooking versus instruction tracing. Techniques like inside-guest kernel-space and outside-guest online can provide the same level of detail but for a system-wide scope [24], [97], [123], [126], [128], [129], [135], [137], [138], [140]. Fine-grained context gives malware analysts the ability to understand the functional capability of a malware behavior (what information malware stores in a file versus what files malware creates). Jiang et al. [118] use network artifacts with fine-grained monitoring in the user-space.

Alternatively, Yin et al. [121] propose *Panorama*, which provides system-wide taint tracking capabilities for observing behavior. The primary goal of *Panorama* is to provide a holistic view of a malware's interaction with the system for *offline* inspection. Other approaches [124], [134], [136], [139], [141], [142] provide a way to track malware interaction through data taint-tracking techniques. Tartarus [139] and FAROS [141] use whole-system emulation to track memory-based attacks, such as cross-process code injection. Although these techniques are promising, they have a large overhead and are difficult to apply for bare-metal sandboxes. Tartarus relies on offline analysis to mitigate the overhead, while FAROS relies on an online approach through record-and-replay with an overhead of 56x.

**Considerations.** Most of the cited work in Table 5 prioritize the transparency of a sandbox deployment. Malware sandbox researchers address transparency through the implementation and monitoring components of a sandbox, but also the analysis environment during a deployment. There are ample techniques [14] to detect virtualized and emulated implementations of malware sandboxes. Randwyk et al. [122] propose a bare-metal-based sandbox to overcome the hardware transparency problem. Several other works [125], [131]–[133], [138] utilize bare-metal sandboxes, but system transparency also affects the environment, which includes OS configuration, installed applications, and network access.

For near-perfect hardware transparency, Nguyen et al. [124] propose using the Secure Virtual Machine (SVM) feature on AMD processors to implement more transparent sandboxes. Jiang et al. [120] show that malware can detect inside-guest monitoring tools and change their behavior. To address monitoring transparency, Jiang et al. propose *VMWatcher*, an outside virtual machine introspection (VMI) technique, to observe malware behavior. However, there is a semantic gap for different sandbox implementations [21].

**Summary.** From a theoretical point of view, sandbox systems balance transparency, isolation, extensibility, and scalability. In practice, they rely on different implementation and monitoring technologies that can create a wide variation in the analysis results due to unforeseen factors. These different configurations have a wide range of applications, which we outline in the systematization. However, the usage has primarily been on understanding the behavior of the binary code of malware. negatively impact sandbox applications.

TABLE 6: A list of installed software in Windows 10 analysis environments.

| Productivity | Browsers | Browser Plugins | Remote Admin | Mail Clients | Instant Messaging |
|---|---|---|---|---|---|
| MS Office 16 | Internet Explorer | Flash 28 | 3D-FTP | OperaMail | ICQ |
| MS Office 2016 | Chrome | Java 8 | CoreFTP | Skype | Trillian |
| Adobe Reader | Firefox | | WinSCP | Thunderbird | Yahoo Messanger |
| | | | FileZilla | FoxMail | WhatsApp |
| | | | NCFTP | Outlook | Pidgin |
| | | | FlashFXP | | |
| | | | BitKinex | | |
| | | | LeechFTP | | |

TABLE 7: An example of signatures used to identify the execution of malware in a sandbox environment.

| Artifact Category | Signature Example |
|---|---|
| File | roaming.bld93115rwr.[a-zA-Z0-9]*\.exe<br>appdata.roaming.[a-z0-9]+.[a-z0-9]+\.exe |
| Registry | software.microsoft.windows.currentversion.run.[a-z]*\.exe<br>downloadmanager.passwords |
| Process (Mutex) | BN[a-f0-9]+<br>[a-zA-Z]{9} |

TABLE 8: Classification accuracy (A), precision (P), recall (R), and F1 score for malware families based on both guided (G) and unguided (U) sandbox artifacts.

| Sandbox Type | Algorithm | All | | | | File | | | | Registry | | | | Process | | | | Network | | | |
|---|---|---|---|---|---|---|---|---|---|---|---|---|---|---|---|---|---|---|---|---|---|
| | | A | P | R | F1 | A | P | R | F1 | A | P | R | F1 | A | P | R | F1 | A | P | R | F1 |
| Guided | Random Forest | 96.52 | 96.77 | 95.3 | 96.57 | 78.26 | 83.77 | 66.37 | 77.9 | 63.91 | 79.54 | 50.08 | 60.71 | 40.43 | 30.55 | 18.92 | 25.68 | 68.26 | 72.74 | 56.91 | 66.96 |
| | Decision Trees | 92.17 | 92.38 | 89.4 | 92.18 | 77.83 | 81.38 | 65.96 | 77.8 | 63.91 | 79.54 | 50.08 | 60.71 | 40.43 | 30.55 | 18.92 | 25.68 | 66.96 | 71.57 | 54.58 | 65.61 |
| | KNN | 90.43 | 90.66 | 88 | 90.41 | 73.48 | 80.14 | 58.49 | 72 | 63.04 | 77.34 | 48.77 | 59.69 | 41.74 | 27.01 | 21.15 | 28.74 | 63.48 | 62.02 | 43.76 | 58.67 |
| | SVM Linear L2 | 85.65 | 87.72 | 82.1 | 85.92 | 64.78 | 57.95 | 43.67 | 59.3 | 45.65 | 31.43 | 25.9 | 32.85 | 33.91 | 12.15 | 13.81 | 17.79 | 44.35 | 50.07 | 22.95 | 37.35 |
| | SVM Dual | 85.65 | 87.72 | 82.1 | 85.92 | 64.78 | 57.95 | 43.67 | 59.3 | 45.65 | 31.43 | 25.9 | 32.85 | 33.91 | 12.15 | 13.81 | 17.79 | 44.35 | 50.07 | 22.95 | 37.35 |
| | Log. Reg. Dual | 84.78 | 87.61 | 79.8 | 85.13 | 67.83 | 65.13 | 50.13 | 64.4 | 46.52 | 36.8 | 27.8 | 34.01 | 37.83 | 14.64 | 13.81 | 21.11 | 43.48 | 54.27 | 22.19 | 36.54 |
| | Log. Reg. L1 | 84.78 | 87.53 | 79.8 | 85.06 | 69.13 | 64.13 | 51.11 | 65.2 | 46.52 | 32.92 | 26.6 | 34.17 | 37.83 | 14.64 | 13.81 | 21.11 | 43.48 | 48.25 | 22.11 | 37.05 |
| | Log. Reg L2 | 84.35 | 87.15 | 78.9 | 84.68 | 67.83 | 65.28 | 50.21 | 64.3 | 45.65 | 38.58 | 27.11 | 32.55 | 37.83 | 14.64 | 13.81 | 21.11 | 43.04 | 54.22 | 21.23 | 36.09 |
| | Preceptron | 83.91 | 84.96 | 78.4 | 83.81 | 50.43 | 60.22 | 28.61 | 46.2 | 23.91 | 11.47 | 20.95 | 13.46 | 10.87 | 4.52 | 2.1 | 4.25 | 37.83 | 52.07 | 30.2 | 36.85 |
| | Preceptron L1 | 79.13 | 81.01 | 74.4 | 79.23 | 59.57 | 65.73 | 44.22 | 55.3 | 47.39 | 54.71 | 33.92 | 39.14 | 11.74 | 3.15 | -4.88 | 4.96 | 41.74 | 61.19 | 36.06 | 40.75 |
| | Preceptron Dual | 72.61 | 71.9 | 66.2 | 69.68 | 47.83 | 72.68 | 30.7 | 49.2 | 43.48 | 24.51 | 24.16 | 28.72 | 10.87 | 4.52 | 2.1 | 4.25 | 23.48 | 43.55 | 15.04 | 20.11 |
| | Preceptron L2 | 60.87 | 72.15 | 47 | 59.34 | 41.74 | 53.27 | 21.82 | 39 | 31.74 | 21.18 | 22.43 | 21.84 | 2.17 | 2.88 | -11.13 | 1.75 | 26.52 | 32.79 | 22.24 | 22.99 |
| | SVM Poly. | 54.78 | 66.31 | 35.6 | 50.02 | 46.09 | 53.22 | 19.56 | 38.6 | 34.35 | 33.84 | 13.87 | 23.22 | 36.96 | 14.75 | 13.33 | 20.94 | 30.87 | 37.21 | 6.67 | 19.32 |
| Unguided | Random Forest | 70.31 | 69.21 | 62.8 | 68.43 | 53.28 | 71.17 | 34.56 | 49.5 | 66.38 | 68.53 | 61.6 | 64.35 | 47.16 | 38.23 | 25.35 | 39.67 | 46.72 | 49.75 | 24.1 | 41.75 |
| | Decision Trees | 67.69 | 67.03 | 62 | 67.01 | 52.84 | 68.54 | 34.22 | 49.2 | 66.81 | 69.5 | 62.02 | 65.03 | 47.16 | 38.23 | 25.35 | 39.67 | 46.72 | 50.52 | 24.32 | 41.71 |
| | KNN | 65.07 | 65.47 | 60.2 | 64.74 | 44.1 | 67.99 | 32.35 | 45.9 | 62.88 | 60.46 | 50.51 | 59.97 | 41.05 | 32.94 | 29.7 | 33.11 | 41.92 | 40.42 | 20.81 | 36.91 |
| | SVM Linear L2 | 64.19 | 63.55 | 47.1 | 59.48 | 48.47 | 51.9 | 27.47 | 41.9 | 58.52 | 57.16 | 37.6 | 52.83 | 25.33 | 7.77 | 0.04 | 11.8 | 39.74 | 55.27 | 15.64 | 32.34 |
| | SVM Dual | 64.19 | 63.55 | 47.1 | 59.48 | 48.47 | 51.9 | 27.47 | 41.9 | 58.52 | 57.16 | 37.6 | 52.83 | 25.33 | 7.77 | 0.04 | 11.8 | 39.74 | 55.27 | 15.64 | 32.34 |
| | Log. Reg. Dual | 62.88 | 62.63 | 45 | 58.01 | 48.47 | 53.71 | 27.47 | 41.9 | 58.95 | 57.65 | 39.22 | 53.68 | 25.33 | 7.77 | 0.04 | 11.8 | 38.43 | 51.3 | 14.34 | 30.61 |
| | Log. Reg. L1 | 62.88 | 62.58 | 45 | 58.04 | 48.03 | 53.95 | 26.57 | 41.2 | 59.39 | 57.74 | 39.39 | 53.79 | 25.33 | 7.77 | 0.04 | 11.8 | 38.86 | 41.02 | 15.53 | 30.73 |
| | Log. Reg L2 | 63.32 | 62.8 | 45.3 | 58.39 | 48.47 | 54.32 | 27.47 | 41.9 | 58.08 | 56.55 | 37.43 | 52.38 | 25.33 | 7.77 | 0.04 | 11.8 | 38.86 | 52.08 | 14.58 | 30.92 |
| | Preceptron | 57.21 | 76.86 | 61 | 55.92 | 43.23 | 48.87 | 23.29 | 37.6 | 43.23 | 63.86 | 33.93 | 38.71 | 17.03 | 3.05 | 13.23 | 5.17 | 27.51 | 34.42 | 17.3 | 26.04 |
| | Preceptron L1 | 63.32 | 58.74 | 45.6 | 57.81 | 43.67 | 54.26 | 23.71 | 38.2 | 40.61 | 62.4 | 31.73 | 36.91 | 25.76 | 6.64 | 0 | 10.56 | 19.21 | 41.83 | 13.59 | 19.93 |
| | Preceptron Dual | 58.08 | 54.64 | 41.2 | 53.12 | 37.99 | 36.01 | 15.33 | 30.8 | 30.57 | 27.32 | 21.53 | 24.03 | 6.55 | 2.98 | -7.77 | 4.04 | 21.83 | 31.26 | 9.53 | 20.56 |
| | Preceptron L2 | 52.84 | 64.23 | 42.1 | 51.28 | 41.48 | 44.47 | 17.28 | 33.5 | 35.37 | 55.53 | 24.62 | 32.78 | 3.93 | 0.87 | -6.98 | 1.42 | 23.58 | 24.04 | 10.56 | 17.57 |
| | SVM Poly. | 46.72 | 63.44 | 26.5 | 42.43 | 41.92 | 55.31 | 21.7 | 36.6 | 48.03 | 56.7 | 28.15 | 42.09 | 33.62 | 19.31 | 7.98 | 22.36 | 32.75 | 25.04 | 5.57 | 21.23 |